\documentclass[reprint,longbibliography,aps,prl]{revtex4-1}
\usepackage[utf8]{inputenc}
\usepackage{amsmath,amssymb,amsthm,mathtools}
\usepackage{graphicx}% Include figure files
\usepackage{dcolumn}% Align table columns on decimal point
\usepackage{bm}% bold math
\usepackage{hyperref}% add hypertext capabilities

\def\be{\begin{equation}}
\def\ee{\end{equation}}
\def\ba#1\ea{\begin{align}#1\end{align}}
\def\bg#1\eg{\begin{gather}#1\end{gather}}
\def\bm#1\em{\begin{multline}#1\end{multline}}
\def\bmd#1\emd{\begin{multlined}#1\end{multlined}}

\def\a{\alpha}
\def\b{\beta}

\def\d{\delta}
\def\D{\Delta}

\def\g{\gamma}
\def\G{\Gamma}

\def\m{\mu}
\def\n{\nu}
\def\p{\phi}

\def\r{\rho}

\def\t{\tau}

\def\W{\Omega}

\def\la{\label}

\def\re{\ref}
\def\er{\eqref}
\def\se{\section}

\def\fr{\frac}
\def\na{\nabla}
\def\pa{\partial}

\def\wtd{\widetilde}

\def\eq{\equiv}

\def\cd{\cdots}

\def\nn{\nonumber}

\def\({\left(}
\def\){\right)}
\def\[{\left[}
\def\]{\right]}
\def\<{\langle}
\def\>{\rangle}

\def\tr{\operatorname{tr}}

\def\bH{{\mathbb H}}
\def\bR{{\mathbb R}}
\def\bZ{{\mathbb Z}}
\def\cA{{\mathcal A}}
\def\cI{{\mathcal I}}
\def\cO{{\mathcal O}}

\def\Area{\operatorname{Area}}
\def\total{\text{total}}
\def\bulk{\text{bulk}}
\def\brane{\text{brane}}
\def\matter{\text{matter}}

\begin{document}

\preprint{1601.06788}
\title{The Gravity Dual of R\'enyi Entropy}
\author{Xi Dong}
\email{xidong@ias.edu}
\affiliation{School of Natural Sciences, Institute for Advanced Study, Princeton, New Jersey 08540, USA}
\date{\today}

\begin{abstract}
A remarkable yet mysterious property of black holes is that their entropy is proportional to the horizon area. This area law inspired the holographic principle, which was later realized concretely in gauge/gravity duality. In this context, entanglement entropy is given by the area of a minimal surface in a dual spacetime. However, discussions of area laws have been constrained to entanglement entropy, whereas a full understanding of a quantum state requires R\'enyi entropies. Here we show that all R\'enyi entropies satisfy a similar area law in holographic theories and are given by the areas of dual cosmic branes. This geometric prescription is a one-parameter generalization of the minimal surface prescription for entanglement entropy. Applying this we provide the first holographic calculation of mutual R\'enyi information between two disks of arbitrary dimension. Our results provide a framework for efficiently studying R\'enyi entropies and understanding entanglement structures in strongly coupled systems and quantum gravity.
\end{abstract}

\maketitle

%%%%%%%%%%%%%%%%%%%%%%%%%%%%%%%%%%%%%%%%%%%%%%%%%%
\se{Introduction}

One of the most remarkable discoveries in fundamental physics is that black holes carry entropy with an amount equal to a quarter of the horizon area in Planck units \cite{Bekenstein:1973ur, Bardeen:1973gs, Hawking:1974sw}:
\be\la{bh}
S = \fr{\Area(\text{Horizon})}{4G_N} \,.
\ee
Here $G_N$ denotes Newton's constant.  The property that gravitational entropies satisfy an area law, however, is not restricted to black holes.  It was elegantly generalized by Ryu and Takayanagi \cite{Ryu:2006bv, Ryu:2006ef} in the context of gauge/gravity duality, an exact equivalence between certain strongly coupled quantum field theories (QFTs) and weakly coupled gravitational theories in one higher dimensions \cite{Maldacena:1997re, Gubser:1998bc, Witten:1998qj}.  In this context, they proposed that the von Neumann entropy, also known as the entanglement entropy, of any spatial region $A$ at a moment of time-reflection symmetry in the boundary QFT is determined by the area of a codimension-2 minimal surface in the dual spacetime:
\be\la{rt}
S = \fr{\Area(\text{Minimal Surface})}{4 G_N} \,.
\ee
The minimal surface is constrained to be at a moment of time-reflection symmetry in the bulk and homologous to the entangling region $A$.  In particular, this means that the minimal surface is anchored at the entangling surface $\pa A$.  Here we follow the standard terminology of referring to the dual spacetime in which the gravitational theory lives as the bulk, and identify the spacetime in which the QFT lives with the asymptotic boundary of the bulk spacetime.

This elegant prescription for holographic entanglement entropy was initially proven in the special case of spherical entangling regions in the vacuum state of a conformal field theory (CFT), by employing a $U(1)$ symmetry to map the problem to one of finding the thermal entropy of the CFT on a hyperboloid \cite{Casini:2011kv}.  The latter problem was then solved by gauge/gravity duality, which tells us that the thermal state of the CFT is dual to a hyperbolic black hole in the bulk, and the thermal entropy is given by the area of the black hole horizon according to Eq.~\er{bh}.  Not surprisingly, this horizon is mapped back to the Ryu--Takayanagi minimal surface in the original problem.

In more general cases, there is no $U(1)$ symmetry to facilitate such a derivation.  Nonetheless, Lewkowycz and Maldacena \cite{Lewkowycz:2013nqa} overcame this difficulty and showed that the Ryu--Takayanagi prescription \er{rt} follows from gauge/gravity duality, by applying the replica trick and generalizing the Euclidean method developed in \cite{Bardeen:1973gs, Gibbons:1976ue, Banados:1993qp, Carlip:1993sa, Susskind:1994sm, Nelson:1994na, Fursaev:1995ef} of calculating gravitational entropies to cases without a $U(1)$ symmetry.  Similar techniques were used in \cite{Dong:2013qoa, Camps:2013zua, Miao:2014nxa} to generalize the Ryu--Takayanagi prescription to cases where the bulk theory involves higher derivative gravity, in \cite{Faulkner:2013ana} to find quantum corrections to the prescription, and in \cite{Dong:2016hjy} to derive a covariant generalization first proposed in \cite{Hubeny:2007xt}.

Discussions of area laws such as Eqs.~\er{bh} and \er{rt} have so far been constrained to the von Neumann entropy.  The main goal of this paper is to generalize these laws to R\'enyi entropies \cite{Renyi:1961, 10.2307/1401301}, which are labeled by an index $n$ and defined in terms of the density matrix $\r$ of the entangling region as
\be\la{renyi}
S_n \eq \fr{1}{1-n} \ln \tr \r^n \,.
\ee
In the $n\to1$ limit we recover the von Neumann entropy $S \eq -\tr(\r\ln\r)$.

Although R\'enyi entropies are often introduced as a one-parameter generalization of the von Neumann entropy, they are much easier to experimentally measure and numerically study (see e.g.\ \cite{Islam:2015measuring} for recent progress in measurements).  They also contain richer physical information about the entanglement structure of a quantum state.  In particular, the knowledge of R\'enyi entropies for all $n$ allows one to determine the whole entanglement spectrum (the set of eigenvalues of $\r$).  R\'enyi entropies have been extensively studied by numerical methods \cite{PhysRevLett.104.157201}, in spin chains \cite{Franchini:2007eu}, in tensor networks \cite{Hayden:2016cfa}, in free field theories \cite{Klebanov:2011uf}, in two-dimensional CFTs \cite{Holzhey:1994we, Lunin:2000yv, Calabrese:2004eu, Calabrese:2009ez, Calabrese:2009qy, Calabrese:2010he, Hartman:2013mia, Chen:2013kpa, Datta:2013hba, Perlmutter:2013paa, Perlmutter:2015iya, Headrick:2015gba} or higher \cite{Perlmutter:2013gua, Lee:2014xwa, Hung:2014npa, Allais:2014ata, Lee:2014zaa, Lewkowycz:2014jia, Bueno:2015rda, Bueno:2015qya, Bueno:2015lza, Bianchi:2015liz, Dong:2016wcf, Bianchi:2016xvf}, and in the context of gauge/gravity duality \cite{Headrick:2010zt, Hung:2011nu, Fursaev:2012mp, Faulkner:2013yia, Galante:2013wta, Belin:2013dva, Barrella:2013wja, Chen:2013dxa}.  They have also been generalized to charged \cite{Belin:2013uta, Belin:2014mva} and supersymmetric cases \cite{Nishioka:2013haa, Alday:2014fsa, Giveon:2015cgs}.

In this paper, we show that all R\'enyi entropies satisfy a similar area law in holographic theories.  As we will see, the gravity dual of R\'enyi entropy is a cosmic brane.  This provides a simple geometric prescription for holographic R\'enyi entropies, generalizing the Ryu--Takayanagi prescription \er{rt} for entanglement entropy.  It is important to distinguish our area law from a universal feature of entanglement and R\'enyi entropies, which is that the most ultraviolet divergent part of these entropies is proportional to the area of the entangling surface in any QFT \cite{Bombelli:1986rw, Srednicki:1993im} (see also \cite{PhysRevB.76.035114, PhysRevLett.100.070502, 1742-5468-2007-08-P08024, Eisert:2008ur} for area-law bounds on entanglement entropy).  Our area law reproduces and goes beyond this universal feature in that it is an exact prescription that applies to both finite and divergent parts of R\'enyi entropies, and it serves as a useful criterion in distinguishing theories with a gravitational dual from those without.

%%%%%%%%%%%%%%%%%%%%%%%%%%%%%%%%%%%%%%%%%%%%%%%%%%
\se{An Area-Law Prescription for R\'enyi Entropy}

Our main result is that a derivative of holographic R\'enyi entropy $S_n$ with respect to the R\'enyi index $n$ satisfies an area law.  It is given by a quarter of the area in Planck units of a bulk codimension-2 cosmic brane homologous to the entangling region:
\be\la{result}
n^2 \pa_n \(\fr{n-1}{n} S_n\) = \fr{\Area(\text{Cosmic Brane}_n)}{4 G_N} \,.
\ee
Here the subscript $n$ on the cosmic brane denotes that its brane tension as a function of $n$ is given by
\be
T_n = \fr{n-1}{4n G_N} \,.
\ee

As shown in Fig.~\re{figal}, the cosmic brane is analogous to the Ryu--Takayanagi minimal surface, except that it backreacts on the ambient geometry by creating a conical deficit angle \cite{Vilenkin:1981zs}
\be\la{coni}
\D\p = 2\pi \fr{n-1}{n} \,.
\ee
\begin{figure}[t]
\centering
\includegraphics[width=\columnwidth]{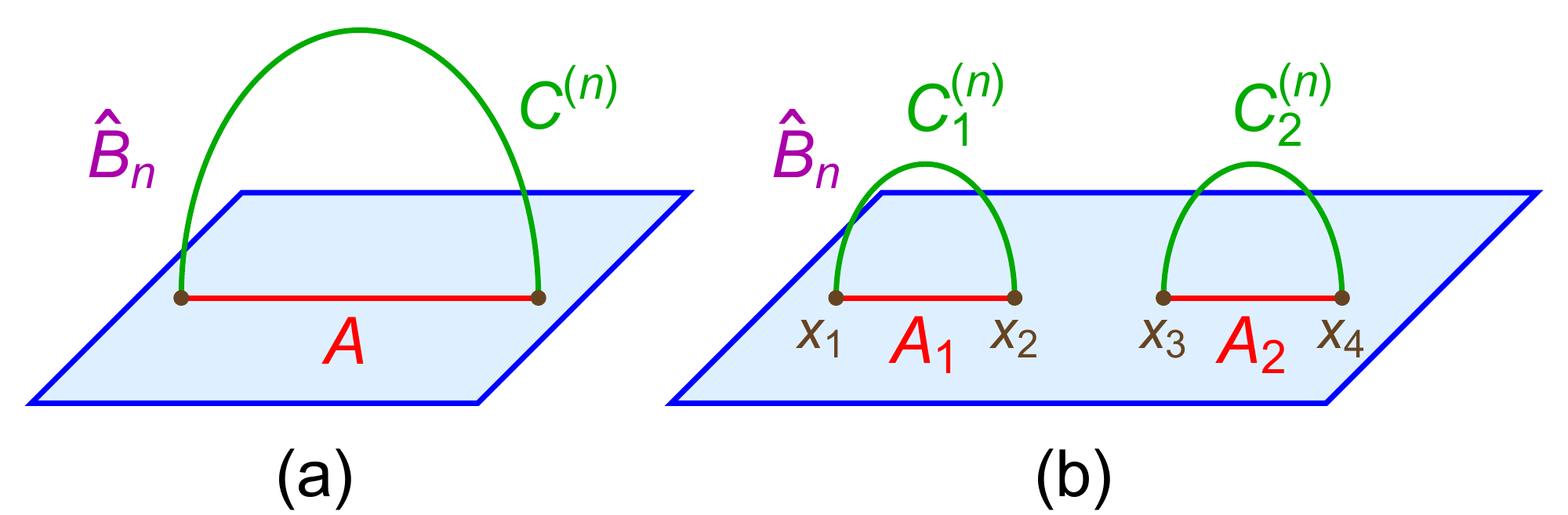}
\caption{Two examples of cosmic branes as the gravity dual of R\'enyi entropy.  The entangling region $A$ (red) is either {\bf (a)} connected or {\bf (b)} disconnected.  In each case a strongly coupled QFT on the plane (blue) has a holographic dual description in terms of a gravitational theory in the bulk spacetime above the plane.  The cosmic brane $C^{(n)}$ (green) is anchored at the entangling surface $\pa A$ (brown) and backreacts on the bulk geometry $\hat B_n$, although the backreaction is difficult to show in the figure.  The R\'enyi entropy is determined by the area of the cosmic brane.  As the R\'enyi index $n$ approaches $1$ the cosmic brane become a non-backreacting minimal surface, reproducing the Ryu--Takayanagi prescription for entanglement entropy.}
\la{figal}
\end{figure}%
A useful way of obtaining the bulk geometry with a cosmic brane is to find the classical solution to the equations of motion resulting from the total (Euclidean) action $I_\total = I_\bulk + I_\brane$, where the bulk action is the Einstein--Hilbert action with matter
\be\la{ibulk}
I_\bulk = -\fr{1}{16\pi G_N} \int d^{d+1}X \sqrt{G} R + I_\matter
\ee
and the brane action is the Nambu--Goto action $I_\brane = T_n \int d^{d-1}y \sqrt{g}.$  Here $X^\m$, $G_{\m\n}$, and $R$ denote the coordinates, metric, and Ricci scalar in the bulk, $y^i$ and $g_{ij}$ denote the coordinates and induced metric on the brane, and $d+1$ is the bulk spacetime dimension.  The action $I_\total$ governs both the bulk metric $G_{\m\n}$ and the embedding map $y^i(X)$ of the brane.  Finding these generally involves solving partial differential equations.  In cases where $I_\total$ admits more than one classical solutions, we choose the dominant solution which has the smallest bulk action $I_\bulk$ (not including $I_\brane$ for reasons that will become clear when we prove our area law).

In the $n\to1$ limit, the cosmic brane becomes tensionless and reduces to a probe brane.  It no longer backreacts on the geometry and instead settles at the location of the minimal surface.  It is manifest that our area law \er{result} reduces to the Ryu--Takayanagi prescription \er{rt} for entanglement entropy in the $n\to1$ limit.  Therefore, our result is a one-parameter generalization of the Ryu--Takayanagi prescription.

Our area law \er{result} suggests that a close variant of the R\'enyi entropy
\be
\wtd{S}_n \eq n^2 \pa_n \(\fr{n-1}{n} S_n\)
\ee
is a more natural candidate for generalizing the von Neumann entropy in that $\wtd{S}_n$ precisely satisfies an area law.  In terms of the density matrix $\r$, it is defined by
\be\la{tsn}
\wtd{S}_n \eq -n^2 \pa_n \(\fr{1}{n} \ln \tr \r^n\) \,.
\ee
It is illuminating to rewrite this as a standard thermodynamic relation
\be\la{tdre}
\wtd{S}_n = -\fr{\pa F_n}{\pa T} \,,
\ee
with the free energy $F_n=-\fr{1}{n} \ln \tr \r^n$ and temperature $T=1/n$.  This can be made precise by noting that $\tr \r^n$ is the canonical partition function with respect to the modular Hamiltonian $K \eq -\ln\r$.  Similar relations between the R\'enyi entropy and free energy have been discussed in \cite{10.1017/CBO9780511524585, Baez:2011}.

In this language, it is useful to compare and contrast our area law \er{result} with the prescription for holographic R\'enyi entropies in \cite{Lewkowycz:2013nqa}.  There a method was developed for calculating the free energy $F_n$ in terms of an on-shell action in the bulk, but our result here goes further in showing that the natural entropy $\wtd{S}_n$ defined by Eq.~\er{tdre} localizes on a codimension-2 surface (the cosmic brane) in the bulk and is determined by the surface area in Planck units.  This localization effect is both conceptually interesting and practically useful.  It allows us to calculate R\'enyi entropies in nontrivial situations as we will see later in an example.  It also agrees with the recent progress in understanding modular Hamiltonians holographically \cite{Jafferis:2014lza, Jafferis:2015del}.

The natural entropy $\wtd{S}_n$ is known to be non-negative for $n>0$ in all quantum systems \cite{10.1017/CBO9780511524585}.  This fact is made manifest by our area law \er{result}.\footnote{We thank E.~Perlmutter for reminding us of this entropy inequality.}  From the quantum information perspective, it is worth studying $\wtd{S}_n$ as a new measure of quantum entanglement.

It is interesting to note that an integrated version of Eq.~\er{result} was used in \cite{Hung:2011nu} to calculate R\'enyi entropies for the special case of spherical entangling regions in a CFT.

It is straightforward to generalize our result \er{result} to theories dual to higher derivative gravity along the directions of \cite{Dong:2013qoa, Camps:2013zua, Miao:2014nxa}.  We replace the right-hand side of Eq.~\er{result} by the Wald entropy \cite{Wald:1993nt, Jacobson:1993vj, Iyer:1994ys} evaluated on a cosmic brane that produces -- in the higher derivative gravity -- a conical deficit angle given by Eq.~\er{coni} for $n \neq 1$, and the extrinsic curvature corrections derived in \cite{Dong:2013qoa, Camps:2013zua, Miao:2014nxa} appear in the $n\to1$ limit.  Similarly, using the methods of \cite{Faulkner:2013ana, Engelhardt:2014gca} we may include quantum corrections to our result in terms of the bulk R\'enyi entropy across the cosmic brane.

%%%%%%%%%%%%%%%%%%%%%%%%%%%%%%%%%%%%%%%%%%%%%%%%%%
\se{Derivation via the Holographic Replica Trick}

We derive the area-law prescription \er{result} for holographic R\'enyi entropies by applying the replica trick in the context of gauge/gravity duality.  We follow the method used in \cite{Lewkowycz:2013nqa} for deriving the Ryu--Takayanagi prescription.

Let us start by recalling that the R\'enyi entropy \er{renyi} of integer index $n>1$ is simply determined by the partition function of the QFT on a branched cover $M_n$, defined by taking $n$ copies of the original Euclidean spacetime $M_1$ on which the QFT lives with a cut along the entangling region and gluing them along the cuts in a cyclic order.  This may be written as
\be\la{snpf}
S_n = \fr{1}{1-n} \(\ln Z[M_n] - n\ln Z[M_1]\) \,,
\ee
where $Z[M_n]$ and $Z[M_1]$ denote the partition function on the branched cover and original spacetime, respectively.  For holographic QFTs, we may calculate $Z[M_n]$ by finding the dominant bulk solution $B_n$ whose asymptotic boundary is $M_n$.  In the so-called large $N$ limit where the bulk physics is classical, we have
\be\la{adscft}
Z[M_n] = e^{-I_\bulk[B_n]}
\ee
where $I_\bulk[B_n]$ denotes the on-shell action of the bulk solution.  Here we work in the Euclidean signature as in \cite{Lewkowycz:2013nqa} by assuming that the entangling region is at a moment of time-reflection symmetry, and leave covariant generalizations along the directions of \cite{Hubeny:2007xt, Dong:2016hjy} to future work.

The branched cover $M_n$ has a manifest $\bZ_n$ symmetry that cyclically permutes the $n$ replicas.  As in \cite{Lewkowycz:2013nqa}, we assume that this $\bZ_n$ replica symmetry is not spontaneously broken in the dominant bulk solution $B_n$.  There is no known example of replica symmetry breaking in the context of gauge/gravity duality although it remains an interesting possibility for further study.  Taking a quotient by the $\bZ_n$ replica symmetry, we define an orbifold
\be\la{orbi}
\hat B_n \eq B_n / \bZ_n \,.
\ee
Since the bulk action is local, we may write
\be\la{slicepizza}
I_\bulk[B_n] = n I_\bulk[\hat B_n] \,.
\ee
Substituting Eqs.~\er{adscft} and \er{slicepizza} into Eq.~\er{snpf}, we find
\be\la{snos}
S_n = \fr{n}{n-1} \(I_\bulk[\hat B_n] - I_\bulk[\hat B_1]\) \,.
\ee

Generally, the branched cover $M_n$ and its bulk dual $B_n$ are meaningful only for integer $n$.  Nonetheless, it is possible to analytically continue the orbifold $\hat B_n$ to non-integer $n$.  The prescription is to use the cosmic brane introduced earlier and find the classical solution resulting from the action $I_\total$.  To see that this gives the same solution as the orbifold \er{orbi} for integer $n$, we note that they have the same codimension-2 conical singularity with the deficit angle \er{coni}.  For the orbifold this results from the regularity of the parent space which is required to satisfy the equations of motion everywhere in the bulk, and the codimension-2 singularity of the orbifold consists of the fixed points of the $\bZ_n$ replica symmetry (see \cite{Haehl:2014zoa} for a detailed discussion of the homology constraint).

Having analytically continued the orbifold $\hat B_n$ to non-integer $n$, we may now use Eq.~\er{snos} to calculate the R\'enyi entropy.  It is worth emphasizing that the on-shell action $I_\bulk[\hat B_n]$ appearing in Eq.~\er{snos} does not include any contribution from the cosmic brane.  This is manifest for integer $n$ because there is no significant contribution from the $\bZ_n$ fixed points to the left-hand side of Eq.~\er{slicepizza}, and the same is true for its right-hand side.  In this sense the cosmic brane is a useful auxiliary tool for generating the relevant conical solutions.

To derive our area law \er{result} for holographic R\'enyi entropies, we note that Eq.~\er{snos} implies
\be\la{dsn}
\pa_n \(\fr{n-1}{n} S_n\) = \pa_n I_\bulk[\hat B_n] \,.
\ee
To evaluate this, we view $\hat B_n$ as a family of classical solutions to the equations of motion with varying boundary conditions at the cosmic brane.  The variation of the on-shell action is therefore a boundary term.  For Einstein gravity \er{ibulk}, we find explicitly
\be\la{bdtm}
\pa_n I_\bulk[\hat B_n] = \int \fr{d^dx \sqrt{\g}}{16\pi G_N} \hat n^\m (\na^\n \pa_n G_{\m\n} - G^{\n\r} \na_\m \pa_n G_{\n\r})
\ee
where we evaluate the integral on a thin codimension-1 tube around the cosmic brane, taking the thickness to zero at the end of the calculation.  Here $x^\a$ and $\g_{\a\b}$ denote the coordinates and induced metric on this tube, and $\hat n^\m$ is the unit normal vector pointing away from the cosmic brane.  The boundary term \er{bdtm} was calculated in the $n\to1$ limit in \cite{Lewkowycz:2013nqa} but here we keep $n$ arbitrary.  We may evaluate Eq.~\er{bdtm} in any coordinate system; one particular choice is to use polar coordinates $r$, $\p$ to describe the 2-dimensional plane orthogonal to the cosmic brane, such that the metric near the brane is
\be
ds^2 = dr^2 +\fr{r^2}{n^2} d\p^2 +g_{ij} dy^i dy^j +\cd \,,
\ee
where the range of $\p$ is fixed as $2\pi$ and $y^i$ denotes the coordinates on the brane.  Evaluating Eq.~\er{bdtm} in this way and using Eq.~\er{dsn}, we find the desired area law \er{result}.

%%%%%%%%%%%%%%%%%%%%%%%%%%%%%%%%%%%%%%%%%%%%%%%%%%
\se{Applications}

Our area law \er{result} gives a simple prescription for calculating holographic R\'enyi entropy of any index $n$ for an arbitrary entangling region.  Applying this prescription is no more difficult than solving partial differential equations, which can be achieved numerically and in some cases analytically.  In principle this allows one to reconstruct the whole entanglement spectrum.  To demonstrate the application of our prescription in a simple example, we provide the first result for the mutual R\'enyi information between two spherical disks in a holographic CFT in $d$ spacetime dimensions.  This is illustrated in Fig.~\re{figal}(b) for the case of $d=2$.

The R\'enyi entropy for a region in a QFT is ultraviolet divergent due to the entanglement between short-distance modes across the entangling surface, but the mutual R\'enyi information between two disjoint regions $A_1$ and $A_2$
\be\la{indef}
I_n(A_1, A_2) \eq S_n(A_1) + S_n(A_2) - S_n(A_1\cup A_2)
\ee
is finite and regulator-independent.  In our example, $A_1$ and $A_2$ represent the two disks.  The mutual R\'enyi information depends on the radii $R_1$, $R_2$ of the disks and the distance $D$ between their centers only through the conformally invariant cross-ratio
\be
x \eq \fr{4 R_1 R_2}{(R_1+D) (R_2+D)} \,.
\ee

Our area law \er{result} reduces the problem to one of finding the bulk solution with a cosmic brane homologous to the union of two disks.  There are two possible topologies for the cosmic brane: either a union of two cosmic branes $C_1^{(n)}$, $C_2^{(n)}$ anchored at $\pa A_1$, $\pa A_2$ respectively as shown in Fig.~\re{figal}(b), or a tube connecting $\pa A_1$ with $\pa A_2$.  We will focus on the first topology which gives the dominant bulk solution for $x<x_c$, with $x_c$ a critical cross-ratio at which the two topologies give equal $I_\bulk$ and a phase transition happens for the R\'enyi entropy.

In the Appendix we derive the mutual R\'enyi information between the two disks to linear order in $\d n\eq n-1$:
\bm\la{inf}
I_n(A_1, A_2) = \fr{2^{3-d} \pi^{d+1} C_T \d n}{d(d^2-1) \G\(\fr{d-1}{2}\)^2} \fr{2-x}{x} \times \\
\times B\(\(\fr{x}{2-x}\)^2; \fr{d+1}{2}, \fr{2-d}{2}\) \,,
\em
where $B$ is the incomplete beta function and $C_T$ is a central charge appearing in the vacuum two-point function of the stress tensor of the CFT in flat space \cite{Osborn:1993cr, Erdmenger:1996yc}:
\be\la{tt}
\<T_{ab}(z) T_{cd}(0)\> = C_T \fr{\cI_{ab,cd}(z)}{z^{2d}} \,.
\ee
Here we have defined $\cI_{ab,cd}(z) \eq \fr{1}{2} [I_{ac}(z) I_{bd}(z) +I_{ad}(z) I_{bc}(z)] - \fr{1}{d} \d_{ab} \d_{cd}$ and $I_{ab}(z) \eq \d_{ab} -2z_a z_b/z^2$.  Our result \er{inf} is derived for CFTs dual to Einstein gravity, but similar techniques can be used for higher derivative gravity (or to higher orders in $\d n$).

In the special case of $d=2$, we may use $C_T = c/2\pi^2$ to rewrite Eq.~\er{inf} in terms of the central charge $c$ of the Virasoro algebra and find at linear order in $\d n$
\be\la{intd}
I_n(A_1, A_2) = \fr{c\d n}{3} \[\fr{x-2}{2x} \ln(1-x) -1\] \,,
\ee
which agrees with the result of a CFT derivation in \cite{Perlmutter:2015iya} and short interval expansions computed in \cite{Headrick:2010zt, Barrella:2013wja, Chen:2013dxa}.

%%%%%%%%%%%%%%%%%%%%%%%%%%%%%%%%%%%%%%%%%%%%%%%%%%
\se{Discussion}

The area law for black holes \er{bh} led to much progress in understanding quantum gravity.  It inspired the holographic principle \cite{'tHooft:1993gx, Susskind:1994vu}, which states that the fundamental degrees of freedom describing any region in quantum gravity are actually encoded on its boundary.  The minimal surface prescription of Ryu and Takayanagi \er{rt} led to numerous results on the von Neumann entropy in strongly coupled systems.  It also played an essential role in establishing a deep and mysterious connection between quantum entanglement and gravity \cite{Jacobson:1995ab, VanRaamsdonk:2010pw, Maldacena:2013xja}.

R\'enyi entropies, however, have until now been more difficult to study than the von Neumann entropy in strongly coupled theories, even though they are more experimentally accessible and contain richer information about a quantum state.  Our area law \er{result} provides a framework for efficiently studying R\'enyi entropies in strongly coupled systems.  Given the recent advances in experimentally measuring R\'enyi entropies, it is worth exploring whether they share similar features with the prediction of our area law in holographic theories.  Progress in this direction would also provide opportunities for distinguishing holographic theories from non-holographic ones, and for eventually understanding the connection between quantum entanglement and gravity.

%%%%%%%%%%%%%%%%%%%%%%%%%%%%%%%%%%%%%%%%%%%%%%%%%%
\se{Acknowledgments}
I thank A.~Lewkowycz, J.~Maldacena, E.~Perlmutter, M.~Rangamani, and A.~Wall for useful discussions, and the Stanford Institute for Theoretical Physics where this work was started.  This work was supported in part by the National Science Foundation under Grant No.~PHY-1316699, by the Department of Energy under Grant No.~DE-SC0009988, and by a Zurich Financial Services Membership at the Institute for Advanced Study.

%%%%%%%%%%%%%%%%%%%%%%%%%%%%%%%%%%%%%%%%%%%%%%%%%%
\appendix
\se{Appendix: Mutual R\'enyi Information Between Two Disks}

We provide the calculation of the R\'enyi entropy for the union of two disjoint disks to linear order in $\d n\eq n-1$, in the vacuum state of a holographic CFT in $d$-dimensional Minkowski space.  Here disks are defined to be perfectly spherical regions in $d-1$ dimensions.

For a single disk, the R\'enyi entropy was calculated in \cite{Hung:2011nu} by using the insight of \cite{Casini:2011kv} to map the problem to one of calculating the thermal entropy of the CFT on a unit hyperboloid at temperature $T=1/2\pi n$.  The holographic calculation then proceeds by finding the relevant hyperbolic black hole in the bulk.  In our language, this means that we know explicitly the bulk solution with a cosmic brane homologous to a single disk.

The mutual R\'enyi information \er{indef} is all that we need to understand the R\'enyi entropy for the union of two disks.  We will express our results in terms of a conformally invariant cross-ratio $x$.  To define $x$ we view the line connecting the two disk centers as a coordinate axis, and it intersects the sphere $\pa A_1$ at $x_1$, $x_2$ and the sphere $\pa A_2$ at $x_3$, $x_4$, as shown in Fig.~\re{figal}(b) for $d=2$.  The cross-ratio $x$ is defined in the standard way from these four points:
\be\la{crdef}
x \eq \fr{(x_1-x_2) (x_3-x_4)}{(x_1-x_3) (x_2-x_4)} \,,
\ee
and it takes real values between 0 and 1.

We will focus on the $x<x_c$ phase in which the dominant bulk solution contains two separate cosmic branes $C_1^{(n)}$, $C_2^{(n)}$ homologous to $A_1$, $A_2$ respectively as shown in Fig.~\re{figal}(b).  Here $x_c$ is a critical cross-ratio where a phase transition happens and the dominant bulk solution for $x>x_c$ switches to one with a tube-like cosmic brane connecting $\pa A_1$ with $\pa A_2$.  For $d=2$, the value of $x_c$ is $1/2$ regardless of $n$, because under the exchange of the entangling region and its complement, we exchange the two brane topologies and $x$ with $1-x$.  In higher dimensions the value of $x_c$ can be numerically calculated for $n=1$ as in \cite{Czech:2014wka} but may generally depend on $n$.

In the $x<x_c$ phase, the mutual information $I_1(A_1,A_2)$ vanishes according to the Ryu--Takayanagi prescription \er{rt}, but for $n\neq1$ the two cosmic branes $C_1^{(n)}$, $C_2^{(n)}$ feel the backreaction of each other.  Applying our area law \er{result} in Eq.~\er{indef} we find
\bg
\pa_n \[\fr{n-1}{n} I_n(A_1, A_2)\] = \fr{\cA(C_1^{(n)}) +\cA(C_2^{(n)}) -\cA(C_{12}^{(n)})}{4 n^2 G_N} \nn\\
= \fr{\cA(C_1^{(n)}) -\cA(C_1^{(n)} | C_2^{(n)}) +\cA(C_2^{(n)}) -\cA(C_2^{(n)} | C_1^{(n)})}{4 n^2 G_N} \,,\la{inr}
\eg
where $\cA$ represents the area and $C_i^{(n)} | C_j^{(n)}$ denotes the cosmic brane $C_i^{(n)}$ in the presence of $C_j^{(n)}$.  Working at linear order in $\d n$, we may replace the area difference $\cA(C_1^{(n)} | C_2^{(n)}) - \cA(C_1^{(n)})$ by
\be\la{adef}
\d\cA_1 \eq \cA(C_1^{(1)} | C_2^{(n)}) - \cA(C_1^{(1)}) \,,
\ee
which is the area variation of the minimal surface $C_1^{(1)}$ homologous to $A_1$ due to the backreaction of $C_2^{(n)}$.  This is because the self-backreaction of the first brane is only affected by the second at higher orders in $\d n$.  Therefore, expanding Eq.~\er{inr} to linear order in $\d n$ we find
\be\la{ins}
I_n(A_1, A_2) = -\fr{\d\cA_1 +\d\cA_2}{8G_N} \,,
\ee
with $\d\cA_2$ defined similarly as in Eq.~\er{adef}.

We will calculate Eq.~\er{ins} in two steps.  First, we find the bulk solution with a cosmic brane homologous to one of the two disks, say the first one $A_1$.  To do this we start by using a suitable conformal transformation to map $A_1$ to the outside of the unit $(d-2)$-sphere and $A_2$ to the inside of a $(d-2)$-sphere with radius $R_0$, with both spheres centered at the origin.  The value of $R_0$ is determined from the cross-ratio by
\be\la{xr}
x = \fr{4R_0}{(1+R_0)^2} \,.
\ee
To see this, we use Eq.~\er{crdef} with $x_1=1$, $x_2=-1$, $x_3=-R_0$, and $x_4=R_0$.

The next step is to conformally map the complement of $A_1$ which is a unit disk to the $(d-1)$-dimensional unit hyperboloid $\bH^{d-1}$.  We do this by noting that the $d$-dimensional flat space metric $ds_{\bR^d}^2 = dt^2 +dR^2 +R^2 d\W_{d-2}^2$ is conformally equivalent to
\be\la{hmet}
ds_{S^1 \times \bH^{d-1}}^2 = d\t^2 + d\r^2 +\sinh^2\r d\W_{d-2}^2 \,,
\ee
which describes $S^1 \times \bH^{d-1}$ with coordinates defined by $\tan\t = 2t/(1-t^2-R^2)$ and $\tanh\r = 2R/(1+t^2+R^2)$.  The $\t$ coordinate has period $2\pi$.  Choosing the two disks to be on $t=0$, we find that the entangling surface of $A_1$ is mapped to $\r=\infty$, and $A_2$ is mapped to the region $\r<\r_0$ with
\be\la{rx}
\r_0 \eq \operatorname{arctanh} \fr{2R_0}{1+R_0^2} = -\fr{1}{2} \ln(1-x) \,,
\ee
where we have used Eq.~\er{xr}.

We may now write down the bulk geometry with a cosmic brane homologous to the first disk $A_1$:
\be\la{hbh}
ds_{d+1}^2 = \fr{dr^2}{f(r)} +f(r) d\t^2 + r^2 \(d\r^2 +\sinh^2\r d\W_{d-2}^2\) \,,
\ee
where for Einstein gravity we have
\be\la{fr}
f(r) = r^2 -1 -\fr{r_h^d -r_h^{d-2}}{r^{d-2}} \,.
\ee
Here we work in the units where the radius of curvature in the asymptotic bulk geometry is set to $1$, and $r$ denotes the holographic direction in the bulk.  As $r\to\infty$ we approach the asymptotic boundary and recover the metric \er{hmet} on $S^1 \times \bH^{d-1}$ conformally.  The bulk \er{hbh} is often viewed as a (Euclidean) black hole geometry with a smooth horizon at $r=r_h$ when $\t$ has period $4\pi/f'(r_h)$, but here we fix the period of $\t$ as $2\pi$ and attribute the conical singularity at $r=r_h$ to a cosmic brane.  Matching the conical deficit angle with Eq.~\er{coni} we find
\be\la{nrh}
n = \fr{2}{f'(r_h)} = \fr{2r_h}{d r_h^2 -d+2} \,.
\ee

We now perform the second step of our calculation which involves finding the minimal surface homologous to the second disk $\r<\r_0$ in the geometry \er{hbh} and calculating its area to linear order in $\d n$.  By spherical symmetry we may describe the minimal surface with a function $r(\r)$.  Its area is
\be\la{afnl}
\cA_2 = S_{d-2} \int_{0}^{\r_0} d\r \sqrt{\fr{r'(\r)^2}{f(r(\r))}+r(\r)^2} \[r(\r) \sinh\r\]^{d-2} \,,
\ee
where $S_{d-2} \eq 2\pi^{(d-1)/2} / \G(\fr{d-1}{2})$ is the area of the unit $(d-2)$-sphere.  The Euler--Lagrange equation from minimizing Eq.~\er{afnl} is difficult to solve in closed form for general $n$, but for $n=1$ the minimal surface is described by
\be\la{rro}
r_{n=1}(\r) = \sqrt{\fr{1-\tanh^2\r}{\tanh^2\r_0 - \tanh^2\r}} \,.
\ee
As we vary $n$ away from $1$, there are two potential contributions to the change of the area \er{afnl}.  The first is due to changing $f(r)$ without moving the surface $r(\r)$, and this contribution may be directly calculated from Eqs.~(\re{fr}-\re{rro}):
\be\la{da}
\d\cA_2 = \fr{\d n S_{d-2} \coth\r_0}{2(1-d)} B\(\tanh^2 \r_0; \fr{d+1}{2}, \fr{2-d}{2}\)
\ee
up to higher orders in $\d n$.  Here $B$ is the incomplete beta function.  The second contribution to the area change is due to moving the surface $r(\r)$ without changing $f(r)$, but to linear order in $\d n$ this is a boundary term at $\r=\r_0$, which may be explicitly shown to vanish.

There is a conformal $\bZ_2$ symmetry that exchanges the two disks, so they are on equal footing and $\d\cA_1 = \d\cA_2$.  Applying Eqs.~\er{rx} and \er{da} in Eq.~\er{ins}, we find that the mutual R\'enyi information between the two disks is
\bm\la{inf2}
I_n(A_1, A_2) = \fr{2^{3-d} \pi^{d+1} C_T \d n}{d(d^2-1) \G\(\fr{d-1}{2}\)^2} \fr{2-x}{x} \times \\
\times B\(\(\fr{x}{2-x}\)^2; \fr{d+1}{2}, \fr{2-d}{2}\) +\cO(\d n^2) \,,
\em
where we have used the relation \cite{Buchel:2009sk}
\be
C_T = \fr{\Gamma(d+2)}{\pi^{d/2} (d-1) \Gamma\(\fr{d}{2}\)} \fr{1}{8 \pi G_N}
\ee
between $G_N$ and the central charge $C_T$ defined in Eq.~\er{tt}.

%%%%%%%%%%%%%%%%%%%%%%%%%%%%%%%%%%%%%%%%%%%%%%%%%%
%\bibliographystyle{apsrev4-1}
\bibliography{bibliography}

\begin{thebibliography}{92}%
\makeatletter
\providecommand \@ifxundefined [1]{%
 \@ifx{#1\undefined}
}%
\providecommand \@ifnum [1]{%
 \ifnum #1\expandafter \@firstoftwo
 \else \expandafter \@secondoftwo
 \fi
}%
\providecommand \@ifx [1]{%
 \ifx #1\expandafter \@firstoftwo
 \else \expandafter \@secondoftwo
 \fi
}%
\providecommand \natexlab [1]{#1}%
\providecommand \enquote  [1]{``#1''}%
\providecommand \bibnamefont  [1]{#1}%
\providecommand \bibfnamefont [1]{#1}%
\providecommand \citenamefont [1]{#1}%
\providecommand \href@noop [0]{\@secondoftwo}%
\providecommand \href [0]{\begingroup \@sanitize@url \@href}%
\providecommand \@href[1]{\@@startlink{#1}\@@href}%
\providecommand \@@href[1]{\endgroup#1\@@endlink}%
\providecommand \@sanitize@url [0]{\catcode `\\12\catcode `\$12\catcode
  `\&12\catcode `\#12\catcode `\^12\catcode `\_12\catcode `\%12\relax}%
\providecommand \@@startlink[1]{}%
\providecommand \@@endlink[0]{}%
\providecommand \url  [0]{\begingroup\@sanitize@url \@url }%
\providecommand \@url [1]{\endgroup\@href {#1}{\urlprefix }}%
\providecommand \urlprefix  [0]{URL }%
\providecommand \Eprint [0]{\href }%
\providecommand \doibase [0]{http://dx.doi.org/}%
\providecommand \selectlanguage [0]{\@gobble}%
\providecommand \bibinfo  [0]{\@secondoftwo}%
\providecommand \bibfield  [0]{\@secondoftwo}%
\providecommand \translation [1]{[#1]}%
\providecommand \BibitemOpen [0]{}%
\providecommand \bibitemStop [0]{}%
\providecommand \bibitemNoStop [0]{.\EOS\space}%
\providecommand \EOS [0]{\spacefactor3000\relax}%
\providecommand \BibitemShut  [1]{\csname bibitem#1\endcsname}%
\let\auto@bib@innerbib\@empty
%</preamble>
\bibitem [{\citenamefont {Bekenstein}(1973)}]{Bekenstein:1973ur}%
  \BibitemOpen
  \bibfield  {author} {\bibinfo {author} {\bibfnamefont {Jacob~D.}\
  \bibnamefont {Bekenstein}},\ }\bibfield  {title} {\enquote {\bibinfo {title}
  {{Black holes and entropy}},}\ }\href {\doibase 10.1103/PhysRevD.7.2333}
  {\bibfield  {journal} {\bibinfo  {journal} {Phys.Rev.}\ }\textbf {\bibinfo
  {volume} {D7}},\ \bibinfo {pages} {2333--2346} (\bibinfo {year}
  {1973})}\BibitemShut {NoStop}%
%%CITATION = PHRVA,D7,2333;%%
\bibitem [{\citenamefont {Bardeen}\ \emph {et~al.}(1973)\citenamefont
  {Bardeen}, \citenamefont {Carter},\ and\ \citenamefont
  {Hawking}}]{Bardeen:1973gs}%
  \BibitemOpen
  \bibfield  {author} {\bibinfo {author} {\bibfnamefont {James~M.}\
  \bibnamefont {Bardeen}}, \bibinfo {author} {\bibfnamefont {B.}~\bibnamefont
  {Carter}}, \ and\ \bibinfo {author} {\bibfnamefont {S.W.}\ \bibnamefont
  {Hawking}},\ }\bibfield  {title} {\enquote {\bibinfo {title} {{The Four laws
  of black hole mechanics}},}\ }\href {\doibase 10.1007/BF01645742} {\bibfield
  {journal} {\bibinfo  {journal} {Commun.Math.Phys.}\ }\textbf {\bibinfo
  {volume} {31}},\ \bibinfo {pages} {161--170} (\bibinfo {year}
  {1973})}\BibitemShut {NoStop}%
%%CITATION = CMPHA,31,161;%%
\bibitem [{\citenamefont {Hawking}(1975)}]{Hawking:1974sw}%
  \BibitemOpen
  \bibfield  {author} {\bibinfo {author} {\bibfnamefont {S.W.}\ \bibnamefont
  {Hawking}},\ }\bibfield  {title} {\enquote {\bibinfo {title} {{Particle
  Creation by Black Holes}},}\ }\href {\doibase 10.1007/BF02345020} {\bibfield
  {journal} {\bibinfo  {journal} {Commun.Math.Phys.}\ }\textbf {\bibinfo
  {volume} {43}},\ \bibinfo {pages} {199--220} (\bibinfo {year}
  {1975})}\BibitemShut {NoStop}%
%%CITATION = CMPHA,43,199;%%
\bibitem [{\citenamefont {Ryu}\ and\ \citenamefont
  {Takayanagi}(2006{\natexlab{a}})}]{Ryu:2006bv}%
  \BibitemOpen
  \bibfield  {author} {\bibinfo {author} {\bibfnamefont {Shinsei}\ \bibnamefont
  {Ryu}}\ and\ \bibinfo {author} {\bibfnamefont {Tadashi}\ \bibnamefont
  {Takayanagi}},\ }\bibfield  {title} {\enquote {\bibinfo {title} {{Holographic
  derivation of entanglement entropy from AdS/CFT}},}\ }\href {\doibase
  10.1103/PhysRevLett.96.181602} {\bibfield  {journal} {\bibinfo  {journal}
  {Phys.Rev.Lett.}\ }\textbf {\bibinfo {volume} {96}},\ \bibinfo {pages}
  {181602} (\bibinfo {year} {2006}{\natexlab{a}})},\ \Eprint
  {http://arxiv.org/abs/hep-th/0603001} {arXiv:hep-th/0603001 [hep-th]}
  \BibitemShut {NoStop}%
%%CITATION = HEP-TH/0603001;%%
\bibitem [{\citenamefont {Ryu}\ and\ \citenamefont
  {Takayanagi}(2006{\natexlab{b}})}]{Ryu:2006ef}%
  \BibitemOpen
  \bibfield  {author} {\bibinfo {author} {\bibfnamefont {Shinsei}\ \bibnamefont
  {Ryu}}\ and\ \bibinfo {author} {\bibfnamefont {Tadashi}\ \bibnamefont
  {Takayanagi}},\ }\bibfield  {title} {\enquote {\bibinfo {title} {{Aspects of
  Holographic Entanglement Entropy}},}\ }\href {\doibase
  10.1088/1126-6708/2006/08/045} {\bibfield  {journal} {\bibinfo  {journal}
  {JHEP}\ }\textbf {\bibinfo {volume} {08}},\ \bibinfo {pages} {045} (\bibinfo
  {year} {2006}{\natexlab{b}})},\ \Eprint {http://arxiv.org/abs/hep-th/0605073}
  {arXiv:hep-th/0605073 [hep-th]} \BibitemShut {NoStop}%
%%CITATION = HEP-TH/0605073;%%
\bibitem [{\citenamefont {Maldacena}(1998)}]{Maldacena:1997re}%
  \BibitemOpen
  \bibfield  {author} {\bibinfo {author} {\bibfnamefont {Juan~Martin}\
  \bibnamefont {Maldacena}},\ }\bibfield  {title} {\enquote {\bibinfo {title}
  {{The Large N limit of superconformal field theories and supergravity}},}\
  }\href@noop {} {\bibfield  {journal} {\bibinfo  {journal}
  {Adv.Theor.Math.Phys.}\ }\textbf {\bibinfo {volume} {2}},\ \bibinfo {pages}
  {231--252} (\bibinfo {year} {1998})},\ \Eprint
  {http://arxiv.org/abs/hep-th/9711200} {arXiv:hep-th/9711200 [hep-th]}
  \BibitemShut {NoStop}%
%%CITATION = HEP-TH/9711200;%%
\bibitem [{\citenamefont {Gubser}\ \emph {et~al.}(1998)\citenamefont {Gubser},
  \citenamefont {Klebanov},\ and\ \citenamefont {Polyakov}}]{Gubser:1998bc}%
  \BibitemOpen
  \bibfield  {author} {\bibinfo {author} {\bibfnamefont {S.S.}\ \bibnamefont
  {Gubser}}, \bibinfo {author} {\bibfnamefont {Igor~R.}\ \bibnamefont
  {Klebanov}}, \ and\ \bibinfo {author} {\bibfnamefont {Alexander~M.}\
  \bibnamefont {Polyakov}},\ }\bibfield  {title} {\enquote {\bibinfo {title}
  {{Gauge theory correlators from noncritical string theory}},}\ }\href
  {\doibase 10.1016/S0370-2693(98)00377-3} {\bibfield  {journal} {\bibinfo
  {journal} {Phys.Lett.}\ }\textbf {\bibinfo {volume} {B428}},\ \bibinfo
  {pages} {105--114} (\bibinfo {year} {1998})},\ \Eprint
  {http://arxiv.org/abs/hep-th/9802109} {arXiv:hep-th/9802109 [hep-th]}
  \BibitemShut {NoStop}%
%%CITATION = HEP-TH/9802109;%%
\bibitem [{\citenamefont {Witten}(1998)}]{Witten:1998qj}%
  \BibitemOpen
  \bibfield  {author} {\bibinfo {author} {\bibfnamefont {Edward}\ \bibnamefont
  {Witten}},\ }\bibfield  {title} {\enquote {\bibinfo {title} {{Anti-de Sitter
  space and holography}},}\ }\href@noop {} {\bibfield  {journal} {\bibinfo
  {journal} {Adv.Theor.Math.Phys.}\ }\textbf {\bibinfo {volume} {2}},\ \bibinfo
  {pages} {253--291} (\bibinfo {year} {1998})},\ \Eprint
  {http://arxiv.org/abs/hep-th/9802150} {arXiv:hep-th/9802150 [hep-th]}
  \BibitemShut {NoStop}%
%%CITATION = HEP-TH/9802150;%%
\bibitem [{\citenamefont {Casini}\ \emph {et~al.}(2011)\citenamefont {Casini},
  \citenamefont {Huerta},\ and\ \citenamefont {Myers}}]{Casini:2011kv}%
  \BibitemOpen
  \bibfield  {author} {\bibinfo {author} {\bibfnamefont {Horacio}\ \bibnamefont
  {Casini}}, \bibinfo {author} {\bibfnamefont {Marina}\ \bibnamefont {Huerta}},
  \ and\ \bibinfo {author} {\bibfnamefont {Robert~C.}\ \bibnamefont {Myers}},\
  }\bibfield  {title} {\enquote {\bibinfo {title} {{Towards a derivation of
  holographic entanglement entropy}},}\ }\href {\doibase
  10.1007/JHEP05(2011)036} {\bibfield  {journal} {\bibinfo  {journal} {JHEP}\
  }\textbf {\bibinfo {volume} {05}},\ \bibinfo {pages} {036} (\bibinfo {year}
  {2011})},\ \Eprint {http://arxiv.org/abs/1102.0440} {arXiv:1102.0440
  [hep-th]} \BibitemShut {NoStop}%
%%CITATION = ARXIV:1102.0440;%%
\bibitem [{\citenamefont {Lewkowycz}\ and\ \citenamefont
  {Maldacena}(2013)}]{Lewkowycz:2013nqa}%
  \BibitemOpen
  \bibfield  {author} {\bibinfo {author} {\bibfnamefont {Aitor}\ \bibnamefont
  {Lewkowycz}}\ and\ \bibinfo {author} {\bibfnamefont {Juan}\ \bibnamefont
  {Maldacena}},\ }\bibfield  {title} {\enquote {\bibinfo {title} {{Generalized
  gravitational entropy}},}\ }\href {\doibase 10.1007/JHEP08(2013)090}
  {\bibfield  {journal} {\bibinfo  {journal} {JHEP}\ }\textbf {\bibinfo
  {volume} {08}},\ \bibinfo {pages} {090} (\bibinfo {year} {2013})},\ \Eprint
  {http://arxiv.org/abs/1304.4926} {arXiv:1304.4926 [hep-th]} \BibitemShut
  {NoStop}%
%%CITATION = ARXIV:1304.4926;%%
\bibitem [{\citenamefont {Gibbons}\ and\ \citenamefont
  {Hawking}(1977)}]{Gibbons:1976ue}%
  \BibitemOpen
  \bibfield  {author} {\bibinfo {author} {\bibfnamefont {G.W.}\ \bibnamefont
  {Gibbons}}\ and\ \bibinfo {author} {\bibfnamefont {S.W.}\ \bibnamefont
  {Hawking}},\ }\bibfield  {title} {\enquote {\bibinfo {title} {{Action
  Integrals and Partition Functions in Quantum Gravity}},}\ }\href {\doibase
  10.1103/PhysRevD.15.2752} {\bibfield  {journal} {\bibinfo  {journal}
  {Phys.Rev.}\ }\textbf {\bibinfo {volume} {D15}},\ \bibinfo {pages}
  {2752--2756} (\bibinfo {year} {1977})}\BibitemShut {NoStop}%
%%CITATION = PHRVA,D15,2752;%%
\bibitem [{\citenamefont {Banados}\ \emph {et~al.}(1994)\citenamefont
  {Banados}, \citenamefont {Teitelboim},\ and\ \citenamefont
  {Zanelli}}]{Banados:1993qp}%
  \BibitemOpen
  \bibfield  {author} {\bibinfo {author} {\bibfnamefont {Maximo}\ \bibnamefont
  {Banados}}, \bibinfo {author} {\bibfnamefont {Claudio}\ \bibnamefont
  {Teitelboim}}, \ and\ \bibinfo {author} {\bibfnamefont {Jorge}\ \bibnamefont
  {Zanelli}},\ }\bibfield  {title} {\enquote {\bibinfo {title} {{Black hole
  entropy and the dimensional continuation of the Gauss-Bonnet theorem}},}\
  }\href {\doibase 10.1103/PhysRevLett.72.957} {\bibfield  {journal} {\bibinfo
  {journal} {Phys.Rev.Lett.}\ }\textbf {\bibinfo {volume} {72}},\ \bibinfo
  {pages} {957--960} (\bibinfo {year} {1994})},\ \Eprint
  {http://arxiv.org/abs/gr-qc/9309026} {arXiv:gr-qc/9309026 [gr-qc]}
  \BibitemShut {NoStop}%
%%CITATION = GR-QC/9309026;%%
\bibitem [{\citenamefont {Carlip}\ and\ \citenamefont
  {Teitelboim}(1995)}]{Carlip:1993sa}%
  \BibitemOpen
  \bibfield  {author} {\bibinfo {author} {\bibfnamefont {Steven}\ \bibnamefont
  {Carlip}}\ and\ \bibinfo {author} {\bibfnamefont {Claudio}\ \bibnamefont
  {Teitelboim}},\ }\bibfield  {title} {\enquote {\bibinfo {title} {{The
  Off-shell black hole}},}\ }\href {\doibase 10.1088/0264-9381/12/7/011}
  {\bibfield  {journal} {\bibinfo  {journal} {Class.Quant.Grav.}\ }\textbf
  {\bibinfo {volume} {12}},\ \bibinfo {pages} {1699--1704} (\bibinfo {year}
  {1995})},\ \Eprint {http://arxiv.org/abs/gr-qc/9312002} {arXiv:gr-qc/9312002
  [gr-qc]} \BibitemShut {NoStop}%
%%CITATION = GR-QC/9312002;%%
\bibitem [{\citenamefont {Susskind}\ and\ \citenamefont
  {Uglum}(1994)}]{Susskind:1994sm}%
  \BibitemOpen
  \bibfield  {author} {\bibinfo {author} {\bibfnamefont {Leonard}\ \bibnamefont
  {Susskind}}\ and\ \bibinfo {author} {\bibfnamefont {John}\ \bibnamefont
  {Uglum}},\ }\bibfield  {title} {\enquote {\bibinfo {title} {{Black hole
  entropy in canonical quantum gravity and superstring theory}},}\ }\href
  {\doibase 10.1103/PhysRevD.50.2700} {\bibfield  {journal} {\bibinfo
  {journal} {Phys.Rev.}\ }\textbf {\bibinfo {volume} {D50}},\ \bibinfo {pages}
  {2700--2711} (\bibinfo {year} {1994})},\ \Eprint
  {http://arxiv.org/abs/hep-th/9401070} {arXiv:hep-th/9401070 [hep-th]}
  \BibitemShut {NoStop}%
%%CITATION = HEP-TH/9401070;%%
\bibitem [{\citenamefont {Nelson}(1994)}]{Nelson:1994na}%
  \BibitemOpen
  \bibfield  {author} {\bibinfo {author} {\bibfnamefont {William}\ \bibnamefont
  {Nelson}},\ }\bibfield  {title} {\enquote {\bibinfo {title} {{A Comment on
  black hole entropy in string theory}},}\ }\href {\doibase
  10.1103/PhysRevD.50.7400} {\bibfield  {journal} {\bibinfo  {journal}
  {Phys.Rev.}\ }\textbf {\bibinfo {volume} {D50}},\ \bibinfo {pages}
  {7400--7402} (\bibinfo {year} {1994})},\ \Eprint
  {http://arxiv.org/abs/hep-th/9406011} {arXiv:hep-th/9406011 [hep-th]}
  \BibitemShut {NoStop}%
%%CITATION = HEP-TH/9406011;%%
\bibitem [{\citenamefont {Fursaev}\ and\ \citenamefont
  {Solodukhin}(1995)}]{Fursaev:1995ef}%
  \BibitemOpen
  \bibfield  {author} {\bibinfo {author} {\bibfnamefont {Dmitri~V.}\
  \bibnamefont {Fursaev}}\ and\ \bibinfo {author} {\bibfnamefont {Sergey~N.}\
  \bibnamefont {Solodukhin}},\ }\bibfield  {title} {\enquote {\bibinfo {title}
  {{On the description of the Riemannian geometry in the presence of conical
  defects}},}\ }\href {\doibase 10.1103/PhysRevD.52.2133} {\bibfield  {journal}
  {\bibinfo  {journal} {Phys.Rev.}\ }\textbf {\bibinfo {volume} {D52}},\
  \bibinfo {pages} {2133--2143} (\bibinfo {year} {1995})},\ \Eprint
  {http://arxiv.org/abs/hep-th/9501127} {arXiv:hep-th/9501127 [hep-th]}
  \BibitemShut {NoStop}%
%%CITATION = HEP-TH/9501127;%%
\bibitem [{\citenamefont {Dong}(2014)}]{Dong:2013qoa}%
  \BibitemOpen
  \bibfield  {author} {\bibinfo {author} {\bibfnamefont {Xi}~\bibnamefont
  {Dong}},\ }\bibfield  {title} {\enquote {\bibinfo {title} {{Holographic
  Entanglement Entropy for General Higher Derivative Gravity}},}\ }\href
  {\doibase 10.1007/JHEP01(2014)044} {\bibfield  {journal} {\bibinfo  {journal}
  {JHEP}\ }\textbf {\bibinfo {volume} {01}},\ \bibinfo {pages} {044} (\bibinfo
  {year} {2014})},\ \Eprint {http://arxiv.org/abs/1310.5713} {arXiv:1310.5713
  [hep-th]} \BibitemShut {NoStop}%
%%CITATION = ARXIV:1310.5713;%%
\bibitem [{\citenamefont {Camps}(2014)}]{Camps:2013zua}%
  \BibitemOpen
  \bibfield  {author} {\bibinfo {author} {\bibfnamefont {Joan}\ \bibnamefont
  {Camps}},\ }\bibfield  {title} {\enquote {\bibinfo {title} {{Generalized
  entropy and higher derivative Gravity}},}\ }\href {\doibase
  10.1007/JHEP03(2014)070} {\bibfield  {journal} {\bibinfo  {journal} {JHEP}\
  }\textbf {\bibinfo {volume} {03}},\ \bibinfo {pages} {070} (\bibinfo {year}
  {2014})},\ \Eprint {http://arxiv.org/abs/1310.6659} {arXiv:1310.6659
  [hep-th]} \BibitemShut {NoStop}%
%%CITATION = ARXIV:1310.6659;%%
\bibitem [{\citenamefont {Miao}\ and\ \citenamefont
  {Guo}(2015)}]{Miao:2014nxa}%
  \BibitemOpen
  \bibfield  {author} {\bibinfo {author} {\bibfnamefont {Rong-Xin}\
  \bibnamefont {Miao}}\ and\ \bibinfo {author} {\bibfnamefont {Wu-zhong}\
  \bibnamefont {Guo}},\ }\bibfield  {title} {\enquote {\bibinfo {title}
  {{Holographic Entanglement Entropy for the Most General Higher Derivative
  Gravity}},}\ }\href {\doibase 10.1007/JHEP08(2015)031} {\bibfield  {journal}
  {\bibinfo  {journal} {JHEP}\ }\textbf {\bibinfo {volume} {08}},\ \bibinfo
  {pages} {031} (\bibinfo {year} {2015})},\ \Eprint
  {http://arxiv.org/abs/1411.5579} {arXiv:1411.5579 [hep-th]} \BibitemShut
  {NoStop}%
%%CITATION = ARXIV:1411.5579;%%
\bibitem [{\citenamefont {Faulkner}\ \emph {et~al.}(2013)\citenamefont
  {Faulkner}, \citenamefont {Lewkowycz},\ and\ \citenamefont
  {Maldacena}}]{Faulkner:2013ana}%
  \BibitemOpen
  \bibfield  {author} {\bibinfo {author} {\bibfnamefont {Thomas}\ \bibnamefont
  {Faulkner}}, \bibinfo {author} {\bibfnamefont {Aitor}\ \bibnamefont
  {Lewkowycz}}, \ and\ \bibinfo {author} {\bibfnamefont {Juan}\ \bibnamefont
  {Maldacena}},\ }\bibfield  {title} {\enquote {\bibinfo {title} {{Quantum
  corrections to holographic entanglement entropy}},}\ }\href {\doibase
  10.1007/JHEP11(2013)074} {\bibfield  {journal} {\bibinfo  {journal} {JHEP}\
  }\textbf {\bibinfo {volume} {11}},\ \bibinfo {pages} {074} (\bibinfo {year}
  {2013})},\ \Eprint {http://arxiv.org/abs/1307.2892} {arXiv:1307.2892
  [hep-th]} \BibitemShut {NoStop}%
%%CITATION = ARXIV:1307.2892;%%
\bibitem [{\citenamefont {Dong}\ \emph {et~al.}(2016)\citenamefont {Dong},
  \citenamefont {Lewkowycz},\ and\ \citenamefont {Rangamani}}]{Dong:2016hjy}%
  \BibitemOpen
  \bibfield  {author} {\bibinfo {author} {\bibfnamefont {Xi}~\bibnamefont
  {Dong}}, \bibinfo {author} {\bibfnamefont {Aitor}\ \bibnamefont {Lewkowycz}},
  \ and\ \bibinfo {author} {\bibfnamefont {Mukund}\ \bibnamefont {Rangamani}},\
  }\bibfield  {title} {\enquote {\bibinfo {title} {{Deriving covariant
  holographic entanglement}},}\ }\href@noop {} {\  (\bibinfo {year} {2016})},\
  \Eprint {http://arxiv.org/abs/1607.07506} {arXiv:1607.07506 [hep-th]}
  \BibitemShut {NoStop}%
%%CITATION = ARXIV:1607.07506;%%
\bibitem [{\citenamefont {Hubeny}\ \emph {et~al.}(2007)\citenamefont {Hubeny},
  \citenamefont {Rangamani},\ and\ \citenamefont {Takayanagi}}]{Hubeny:2007xt}%
  \BibitemOpen
  \bibfield  {author} {\bibinfo {author} {\bibfnamefont {Veronika~E.}\
  \bibnamefont {Hubeny}}, \bibinfo {author} {\bibfnamefont {Mukund}\
  \bibnamefont {Rangamani}}, \ and\ \bibinfo {author} {\bibfnamefont {Tadashi}\
  \bibnamefont {Takayanagi}},\ }\bibfield  {title} {\enquote {\bibinfo {title}
  {{A Covariant holographic entanglement entropy proposal}},}\ }\href {\doibase
  10.1088/1126-6708/2007/07/062} {\bibfield  {journal} {\bibinfo  {journal}
  {JHEP}\ }\textbf {\bibinfo {volume} {07}},\ \bibinfo {pages} {062} (\bibinfo
  {year} {2007})},\ \Eprint {http://arxiv.org/abs/0705.0016} {arXiv:0705.0016
  [hep-th]} \BibitemShut {NoStop}%
%%CITATION = ARXIV:0705.0016;%%
\bibitem [{\citenamefont {Rényi}(1961)}]{Renyi:1961}%
  \BibitemOpen
  \bibfield  {author} {\bibinfo {author} {\bibfnamefont {Alfréd}\ \bibnamefont
  {Rényi}},\ }\bibfield  {title} {\enquote {\bibinfo {title} {On measures of
  entropy and information},}\ }in\ \href
  {http://projecteuclid.org/euclid.bsmsp/1200512181} {\emph {\bibinfo
  {booktitle} {Proceedings of the Fourth Berkeley Symposium on Mathematical
  Statistics and Probability, Volume 1: Contributions to the Theory of
  Statistics}}}\ (\bibinfo  {publisher} {University of California Press},\
  \bibinfo {year} {1961})\ pp.\ \bibinfo {pages} {547--561}\BibitemShut
  {NoStop}%
\bibitem [{\citenamefont {Rényi}(1965)}]{10.2307/1401301}%
  \BibitemOpen
  \bibfield  {author} {\bibinfo {author} {\bibfnamefont {A.}~\bibnamefont
  {Rényi}},\ }\bibfield  {title} {\enquote {\bibinfo {title} {On the
  foundations of information theory},}\ }\href
  {http://www.jstor.org/stable/1401301} {\bibfield  {journal} {\bibinfo
  {journal} {Revue de l'Institut International de Statistique / Review of the
  International Statistical Institute}\ }\textbf {\bibinfo {volume} {33}},\
  \bibinfo {pages} {1--14} (\bibinfo {year} {1965})}\BibitemShut {NoStop}%
\bibitem [{\citenamefont {Islam}\ \emph {et~al.}(2015)\citenamefont {Islam},
  \citenamefont {Ma}, \citenamefont {Preiss}, \citenamefont {Tai},
  \citenamefont {Lukin}, \citenamefont {Rispoli},\ and\ \citenamefont
  {Greiner}}]{Islam:2015measuring}%
  \BibitemOpen
  \bibfield  {author} {\bibinfo {author} {\bibfnamefont {Rajibul}\ \bibnamefont
  {Islam}}, \bibinfo {author} {\bibfnamefont {Ruichao}\ \bibnamefont {Ma}},
  \bibinfo {author} {\bibfnamefont {Philipp~M.}\ \bibnamefont {Preiss}},
  \bibinfo {author} {\bibfnamefont {M.~Eric}\ \bibnamefont {Tai}}, \bibinfo
  {author} {\bibfnamefont {Alexander}\ \bibnamefont {Lukin}}, \bibinfo {author}
  {\bibfnamefont {Matthew}\ \bibnamefont {Rispoli}}, \ and\ \bibinfo {author}
  {\bibfnamefont {Markus}\ \bibnamefont {Greiner}},\ }\bibfield  {title}
  {\enquote {\bibinfo {title} {Measuring entanglement entropy in a quantum
  many-body system},}\ }\href@noop {} {\bibfield  {journal} {\bibinfo
  {journal} {Nature}\ }\textbf {\bibinfo {volume} {528}},\ \bibinfo {pages}
  {77--83} (\bibinfo {year} {2015})}\BibitemShut {NoStop}%
\bibitem [{\citenamefont {Hastings}\ \emph {et~al.}(2010)\citenamefont
  {Hastings}, \citenamefont {Gonz\'alez}, \citenamefont {Kallin},\ and\
  \citenamefont {Melko}}]{PhysRevLett.104.157201}%
  \BibitemOpen
  \bibfield  {author} {\bibinfo {author} {\bibfnamefont {Matthew~B.}\
  \bibnamefont {Hastings}}, \bibinfo {author} {\bibfnamefont {Iv\'an}\
  \bibnamefont {Gonz\'alez}}, \bibinfo {author} {\bibfnamefont {Ann~B.}\
  \bibnamefont {Kallin}}, \ and\ \bibinfo {author} {\bibfnamefont {Roger~G.}\
  \bibnamefont {Melko}},\ }\bibfield  {title} {\enquote {\bibinfo {title}
  {Measuring renyi entanglement entropy in quantum monte carlo simulations},}\
  }\href {\doibase 10.1103/PhysRevLett.104.157201} {\bibfield  {journal}
  {\bibinfo  {journal} {Phys. Rev. Lett.}\ }\textbf {\bibinfo {volume} {104}},\
  \bibinfo {pages} {157201} (\bibinfo {year} {2010})}\BibitemShut {NoStop}%
\bibitem [{\citenamefont {Franchini}\ \emph {et~al.}(2008)\citenamefont
  {Franchini}, \citenamefont {Its},\ and\ \citenamefont
  {Korepin}}]{Franchini:2007eu}%
  \BibitemOpen
  \bibfield  {author} {\bibinfo {author} {\bibfnamefont {F.}~\bibnamefont
  {Franchini}}, \bibinfo {author} {\bibfnamefont {A.~R.}\ \bibnamefont {Its}},
  \ and\ \bibinfo {author} {\bibfnamefont {V.~E.}\ \bibnamefont {Korepin}},\
  }\bibfield  {title} {\enquote {\bibinfo {title} {{Renyi Entropy of the XY
  Spin Chain}},}\ }\href {\doibase 10.1088/1751-8113/41/2/025302} {\bibfield
  {journal} {\bibinfo  {journal} {J. Phys.}\ }\textbf {\bibinfo {volume}
  {A41}},\ \bibinfo {pages} {025302} (\bibinfo {year} {2008})},\ \Eprint
  {http://arxiv.org/abs/0707.2534} {arXiv:0707.2534 [quant-ph]} \BibitemShut
  {NoStop}%
%%CITATION = ARXIV:0707.2534;%%
\bibitem [{\citenamefont {Hayden}\ \emph {et~al.}(2016)\citenamefont {Hayden},
  \citenamefont {Nezami}, \citenamefont {Qi}, \citenamefont {Thomas},
  \citenamefont {Walter},\ and\ \citenamefont {Yang}}]{Hayden:2016cfa}%
  \BibitemOpen
  \bibfield  {author} {\bibinfo {author} {\bibfnamefont {Patrick}\ \bibnamefont
  {Hayden}}, \bibinfo {author} {\bibfnamefont {Sepehr}\ \bibnamefont {Nezami}},
  \bibinfo {author} {\bibfnamefont {Xiao-Liang}\ \bibnamefont {Qi}}, \bibinfo
  {author} {\bibfnamefont {Nathaniel}\ \bibnamefont {Thomas}}, \bibinfo
  {author} {\bibfnamefont {Michael}\ \bibnamefont {Walter}}, \ and\ \bibinfo
  {author} {\bibfnamefont {Zhao}\ \bibnamefont {Yang}},\ }\bibfield  {title}
  {\enquote {\bibinfo {title} {{Holographic duality from random tensor
  networks}},}\ }\href@noop {} {\  (\bibinfo {year} {2016})},\ \Eprint
  {http://arxiv.org/abs/1601.01694} {arXiv:1601.01694 [hep-th]} \BibitemShut
  {NoStop}%
%%CITATION = ARXIV:1601.01694;%%
\bibitem [{\citenamefont {Klebanov}\ \emph {et~al.}(2012)\citenamefont
  {Klebanov}, \citenamefont {Pufu}, \citenamefont {Sachdev},\ and\
  \citenamefont {Safdi}}]{Klebanov:2011uf}%
  \BibitemOpen
  \bibfield  {author} {\bibinfo {author} {\bibfnamefont {Igor~R.}\ \bibnamefont
  {Klebanov}}, \bibinfo {author} {\bibfnamefont {Silviu~S.}\ \bibnamefont
  {Pufu}}, \bibinfo {author} {\bibfnamefont {Subir}\ \bibnamefont {Sachdev}}, \
  and\ \bibinfo {author} {\bibfnamefont {Benjamin~R.}\ \bibnamefont {Safdi}},\
  }\bibfield  {title} {\enquote {\bibinfo {title} {{Renyi Entropies for Free
  Field Theories}},}\ }\href {\doibase 10.1007/JHEP04(2012)074} {\bibfield
  {journal} {\bibinfo  {journal} {JHEP}\ }\textbf {\bibinfo {volume} {04}},\
  \bibinfo {pages} {074} (\bibinfo {year} {2012})},\ \Eprint
  {http://arxiv.org/abs/1111.6290} {arXiv:1111.6290 [hep-th]} \BibitemShut
  {NoStop}%
%%CITATION = ARXIV:1111.6290;%%
\bibitem [{\citenamefont {Holzhey}\ \emph {et~al.}(1994)\citenamefont
  {Holzhey}, \citenamefont {Larsen},\ and\ \citenamefont
  {Wilczek}}]{Holzhey:1994we}%
  \BibitemOpen
  \bibfield  {author} {\bibinfo {author} {\bibfnamefont {Christoph}\
  \bibnamefont {Holzhey}}, \bibinfo {author} {\bibfnamefont {Finn}\
  \bibnamefont {Larsen}}, \ and\ \bibinfo {author} {\bibfnamefont {Frank}\
  \bibnamefont {Wilczek}},\ }\bibfield  {title} {\enquote {\bibinfo {title}
  {{Geometric and renormalized entropy in conformal field theory}},}\ }\href
  {\doibase 10.1016/0550-3213(94)90402-2} {\bibfield  {journal} {\bibinfo
  {journal} {Nucl. Phys.}\ }\textbf {\bibinfo {volume} {B424}},\ \bibinfo
  {pages} {443--467} (\bibinfo {year} {1994})},\ \Eprint
  {http://arxiv.org/abs/hep-th/9403108} {arXiv:hep-th/9403108 [hep-th]}
  \BibitemShut {NoStop}%
%%CITATION = HEP-TH/9403108;%%
\bibitem [{\citenamefont {Lunin}\ and\ \citenamefont
  {Mathur}(2001)}]{Lunin:2000yv}%
  \BibitemOpen
  \bibfield  {author} {\bibinfo {author} {\bibfnamefont {Oleg}\ \bibnamefont
  {Lunin}}\ and\ \bibinfo {author} {\bibfnamefont {Samir~D.}\ \bibnamefont
  {Mathur}},\ }\bibfield  {title} {\enquote {\bibinfo {title} {{Correlation
  functions for M**N / S(N) orbifolds}},}\ }\href {\doibase
  10.1007/s002200100431} {\bibfield  {journal} {\bibinfo  {journal} {Commun.
  Math. Phys.}\ }\textbf {\bibinfo {volume} {219}},\ \bibinfo {pages}
  {399--442} (\bibinfo {year} {2001})},\ \Eprint
  {http://arxiv.org/abs/hep-th/0006196} {arXiv:hep-th/0006196 [hep-th]}
  \BibitemShut {NoStop}%
%%CITATION = HEP-TH/0006196;%%
\bibitem [{\citenamefont {Calabrese}\ and\ \citenamefont
  {Cardy}(2004)}]{Calabrese:2004eu}%
  \BibitemOpen
  \bibfield  {author} {\bibinfo {author} {\bibfnamefont {Pasquale}\
  \bibnamefont {Calabrese}}\ and\ \bibinfo {author} {\bibfnamefont {John~L.}\
  \bibnamefont {Cardy}},\ }\bibfield  {title} {\enquote {\bibinfo {title}
  {{Entanglement entropy and quantum field theory}},}\ }\href {\doibase
  10.1088/1742-5468/2004/06/P06002} {\bibfield  {journal} {\bibinfo  {journal}
  {J. Stat. Mech.}\ }\textbf {\bibinfo {volume} {0406}},\ \bibinfo {pages}
  {P06002} (\bibinfo {year} {2004})},\ \Eprint
  {http://arxiv.org/abs/hep-th/0405152} {arXiv:hep-th/0405152 [hep-th]}
  \BibitemShut {NoStop}%
%%CITATION = HEP-TH/0405152;%%
\bibitem [{\citenamefont {Calabrese}\ \emph {et~al.}(2009)\citenamefont
  {Calabrese}, \citenamefont {Cardy},\ and\ \citenamefont
  {Tonni}}]{Calabrese:2009ez}%
  \BibitemOpen
  \bibfield  {author} {\bibinfo {author} {\bibfnamefont {Pasquale}\
  \bibnamefont {Calabrese}}, \bibinfo {author} {\bibfnamefont {John}\
  \bibnamefont {Cardy}}, \ and\ \bibinfo {author} {\bibfnamefont {Erik}\
  \bibnamefont {Tonni}},\ }\bibfield  {title} {\enquote {\bibinfo {title}
  {{Entanglement entropy of two disjoint intervals in conformal field
  theory}},}\ }\href {\doibase 10.1088/1742-5468/2009/11/P11001} {\bibfield
  {journal} {\bibinfo  {journal} {J. Stat. Mech.}\ }\textbf {\bibinfo {volume}
  {0911}},\ \bibinfo {pages} {P11001} (\bibinfo {year} {2009})},\ \Eprint
  {http://arxiv.org/abs/0905.2069} {arXiv:0905.2069 [hep-th]} \BibitemShut
  {NoStop}%
%%CITATION = ARXIV:0905.2069;%%
\bibitem [{\citenamefont {Calabrese}\ and\ \citenamefont
  {Cardy}(2009)}]{Calabrese:2009qy}%
  \BibitemOpen
  \bibfield  {author} {\bibinfo {author} {\bibfnamefont {Pasquale}\
  \bibnamefont {Calabrese}}\ and\ \bibinfo {author} {\bibfnamefont {John}\
  \bibnamefont {Cardy}},\ }\bibfield  {title} {\enquote {\bibinfo {title}
  {{Entanglement entropy and conformal field theory}},}\ }\href {\doibase
  10.1088/1751-8113/42/50/504005} {\bibfield  {journal} {\bibinfo  {journal}
  {J. Phys.}\ }\textbf {\bibinfo {volume} {A42}},\ \bibinfo {pages} {504005}
  (\bibinfo {year} {2009})},\ \Eprint {http://arxiv.org/abs/0905.4013}
  {arXiv:0905.4013 [cond-mat.stat-mech]} \BibitemShut {NoStop}%
%%CITATION = ARXIV:0905.4013;%%
\bibitem [{\citenamefont {Calabrese}\ \emph {et~al.}(2011)\citenamefont
  {Calabrese}, \citenamefont {Cardy},\ and\ \citenamefont
  {Tonni}}]{Calabrese:2010he}%
  \BibitemOpen
  \bibfield  {author} {\bibinfo {author} {\bibfnamefont {Pasquale}\
  \bibnamefont {Calabrese}}, \bibinfo {author} {\bibfnamefont {John}\
  \bibnamefont {Cardy}}, \ and\ \bibinfo {author} {\bibfnamefont {Erik}\
  \bibnamefont {Tonni}},\ }\bibfield  {title} {\enquote {\bibinfo {title}
  {{Entanglement entropy of two disjoint intervals in conformal field theory
  II}},}\ }\href {\doibase 10.1088/1742-5468/2011/01/P01021} {\bibfield
  {journal} {\bibinfo  {journal} {J. Stat. Mech.}\ }\textbf {\bibinfo {volume}
  {1101}},\ \bibinfo {pages} {P01021} (\bibinfo {year} {2011})},\ \Eprint
  {http://arxiv.org/abs/1011.5482} {arXiv:1011.5482 [hep-th]} \BibitemShut
  {NoStop}%
%%CITATION = ARXIV:1011.5482;%%
\bibitem [{\citenamefont {Hartman}(2013)}]{Hartman:2013mia}%
  \BibitemOpen
  \bibfield  {author} {\bibinfo {author} {\bibfnamefont {Thomas}\ \bibnamefont
  {Hartman}},\ }\bibfield  {title} {\enquote {\bibinfo {title} {{Entanglement
  Entropy at Large Central Charge}},}\ }\href@noop {} {\  (\bibinfo {year}
  {2013})},\ \Eprint {http://arxiv.org/abs/1303.6955} {arXiv:1303.6955
  [hep-th]} \BibitemShut {NoStop}%
%%CITATION = ARXIV:1303.6955;%%
\bibitem [{\citenamefont {Chen}\ and\ \citenamefont
  {Zhang}(2013)}]{Chen:2013kpa}%
  \BibitemOpen
  \bibfield  {author} {\bibinfo {author} {\bibfnamefont {Bin}\ \bibnamefont
  {Chen}}\ and\ \bibinfo {author} {\bibfnamefont {Jia-Ju}\ \bibnamefont
  {Zhang}},\ }\bibfield  {title} {\enquote {\bibinfo {title} {{On short
  interval expansion of Rényi entropy}},}\ }\href {\doibase
  10.1007/JHEP11(2013)164} {\bibfield  {journal} {\bibinfo  {journal} {JHEP}\
  }\textbf {\bibinfo {volume} {11}},\ \bibinfo {pages} {164} (\bibinfo {year}
  {2013})},\ \Eprint {http://arxiv.org/abs/1309.5453} {arXiv:1309.5453
  [hep-th]} \BibitemShut {NoStop}%
%%CITATION = ARXIV:1309.5453;%%
\bibitem [{\citenamefont {Datta}\ and\ \citenamefont
  {David}(2014)}]{Datta:2013hba}%
  \BibitemOpen
  \bibfield  {author} {\bibinfo {author} {\bibfnamefont {Shouvik}\ \bibnamefont
  {Datta}}\ and\ \bibinfo {author} {\bibfnamefont {Justin~R.}\ \bibnamefont
  {David}},\ }\bibfield  {title} {\enquote {\bibinfo {title} {{Rényi entropies
  of free bosons on the torus and holography}},}\ }\href {\doibase
  10.1007/JHEP04(2014)081} {\bibfield  {journal} {\bibinfo  {journal} {JHEP}\
  }\textbf {\bibinfo {volume} {04}},\ \bibinfo {pages} {081} (\bibinfo {year}
  {2014})},\ \Eprint {http://arxiv.org/abs/1311.1218} {arXiv:1311.1218
  [hep-th]} \BibitemShut {NoStop}%
%%CITATION = ARXIV:1311.1218;%%
\bibitem [{\citenamefont
  {Perlmutter}(2014{\natexlab{a}})}]{Perlmutter:2013paa}%
  \BibitemOpen
  \bibfield  {author} {\bibinfo {author} {\bibfnamefont {Eric}\ \bibnamefont
  {Perlmutter}},\ }\bibfield  {title} {\enquote {\bibinfo {title} {{Comments on
  Renyi entropy in AdS$_3$/CFT$_2$}},}\ }\href {\doibase
  10.1007/JHEP05(2014)052} {\bibfield  {journal} {\bibinfo  {journal} {JHEP}\
  }\textbf {\bibinfo {volume} {05}},\ \bibinfo {pages} {052} (\bibinfo {year}
  {2014}{\natexlab{a}})},\ \Eprint {http://arxiv.org/abs/1312.5740}
  {arXiv:1312.5740 [hep-th]} \BibitemShut {NoStop}%
%%CITATION = ARXIV:1312.5740;%%
\bibitem [{\citenamefont {Perlmutter}(2015)}]{Perlmutter:2015iya}%
  \BibitemOpen
  \bibfield  {author} {\bibinfo {author} {\bibfnamefont {Eric}\ \bibnamefont
  {Perlmutter}},\ }\bibfield  {title} {\enquote {\bibinfo {title} {{Virasoro
  conformal blocks in closed form}},}\ }\href {\doibase
  10.1007/JHEP08(2015)088} {\bibfield  {journal} {\bibinfo  {journal} {JHEP}\
  }\textbf {\bibinfo {volume} {08}},\ \bibinfo {pages} {088} (\bibinfo {year}
  {2015})},\ \Eprint {http://arxiv.org/abs/1502.07742} {arXiv:1502.07742
  [hep-th]} \BibitemShut {NoStop}%
%%CITATION = ARXIV:1502.07742;%%
\bibitem [{\citenamefont {Headrick}\ \emph {et~al.}(2015)\citenamefont
  {Headrick}, \citenamefont {Maloney}, \citenamefont {Perlmutter},\ and\
  \citenamefont {Zadeh}}]{Headrick:2015gba}%
  \BibitemOpen
  \bibfield  {author} {\bibinfo {author} {\bibfnamefont {Matthew}\ \bibnamefont
  {Headrick}}, \bibinfo {author} {\bibfnamefont {Alexander}\ \bibnamefont
  {Maloney}}, \bibinfo {author} {\bibfnamefont {Eric}\ \bibnamefont
  {Perlmutter}}, \ and\ \bibinfo {author} {\bibfnamefont {Ida~G.}\ \bibnamefont
  {Zadeh}},\ }\bibfield  {title} {\enquote {\bibinfo {title} {{Rényi
  entropies, the analytic bootstrap, and 3D quantum gravity at higher
  genus}},}\ }\href {\doibase 10.1007/JHEP07(2015)059} {\bibfield  {journal}
  {\bibinfo  {journal} {JHEP}\ }\textbf {\bibinfo {volume} {07}},\ \bibinfo
  {pages} {059} (\bibinfo {year} {2015})},\ \Eprint
  {http://arxiv.org/abs/1503.07111} {arXiv:1503.07111 [hep-th]} \BibitemShut
  {NoStop}%
%%CITATION = ARXIV:1503.07111;%%
\bibitem [{\citenamefont
  {Perlmutter}(2014{\natexlab{b}})}]{Perlmutter:2013gua}%
  \BibitemOpen
  \bibfield  {author} {\bibinfo {author} {\bibfnamefont {Eric}\ \bibnamefont
  {Perlmutter}},\ }\bibfield  {title} {\enquote {\bibinfo {title} {{A universal
  feature of CFT Rényi entropy}},}\ }\href {\doibase 10.1007/JHEP03(2014)117}
  {\bibfield  {journal} {\bibinfo  {journal} {JHEP}\ }\textbf {\bibinfo
  {volume} {03}},\ \bibinfo {pages} {117} (\bibinfo {year}
  {2014}{\natexlab{b}})},\ \Eprint {http://arxiv.org/abs/1308.1083}
  {arXiv:1308.1083 [hep-th]} \BibitemShut {NoStop}%
%%CITATION = ARXIV:1308.1083;%%
\bibitem [{\citenamefont {Lee}\ \emph {et~al.}(2014)\citenamefont {Lee},
  \citenamefont {McGough},\ and\ \citenamefont {Safdi}}]{Lee:2014xwa}%
  \BibitemOpen
  \bibfield  {author} {\bibinfo {author} {\bibfnamefont {Jeongseog}\
  \bibnamefont {Lee}}, \bibinfo {author} {\bibfnamefont {Lauren}\ \bibnamefont
  {McGough}}, \ and\ \bibinfo {author} {\bibfnamefont {Benjamin~R.}\
  \bibnamefont {Safdi}},\ }\bibfield  {title} {\enquote {\bibinfo {title}
  {{Rényi entropy and geometry}},}\ }\href {\doibase
  10.1103/PhysRevD.89.125016} {\bibfield  {journal} {\bibinfo  {journal} {Phys.
  Rev.}\ }\textbf {\bibinfo {volume} {D89}},\ \bibinfo {pages} {125016}
  (\bibinfo {year} {2014})},\ \Eprint {http://arxiv.org/abs/1403.1580}
  {arXiv:1403.1580 [hep-th]} \BibitemShut {NoStop}%
%%CITATION = ARXIV:1403.1580;%%
\bibitem [{\citenamefont {Hung}\ \emph {et~al.}(2014)\citenamefont {Hung},
  \citenamefont {Myers},\ and\ \citenamefont {Smolkin}}]{Hung:2014npa}%
  \BibitemOpen
  \bibfield  {author} {\bibinfo {author} {\bibfnamefont {Ling-Yan}\
  \bibnamefont {Hung}}, \bibinfo {author} {\bibfnamefont {Robert~C.}\
  \bibnamefont {Myers}}, \ and\ \bibinfo {author} {\bibfnamefont {Michael}\
  \bibnamefont {Smolkin}},\ }\bibfield  {title} {\enquote {\bibinfo {title}
  {{Twist operators in higher dimensions}},}\ }\href {\doibase
  10.1007/JHEP10(2014)178} {\bibfield  {journal} {\bibinfo  {journal} {JHEP}\
  }\textbf {\bibinfo {volume} {10}},\ \bibinfo {pages} {178} (\bibinfo {year}
  {2014})},\ \Eprint {http://arxiv.org/abs/1407.6429} {arXiv:1407.6429
  [hep-th]} \BibitemShut {NoStop}%
%%CITATION = ARXIV:1407.6429;%%
\bibitem [{\citenamefont {Allais}\ and\ \citenamefont
  {Mezei}(2015)}]{Allais:2014ata}%
  \BibitemOpen
  \bibfield  {author} {\bibinfo {author} {\bibfnamefont {Andrea}\ \bibnamefont
  {Allais}}\ and\ \bibinfo {author} {\bibfnamefont {Márk}\ \bibnamefont
  {Mezei}},\ }\bibfield  {title} {\enquote {\bibinfo {title} {{Some results on
  the shape dependence of entanglement and Rényi entropies}},}\ }\href
  {\doibase 10.1103/PhysRevD.91.046002} {\bibfield  {journal} {\bibinfo
  {journal} {Phys. Rev.}\ }\textbf {\bibinfo {volume} {D91}},\ \bibinfo {pages}
  {046002} (\bibinfo {year} {2015})},\ \Eprint {http://arxiv.org/abs/1407.7249}
  {arXiv:1407.7249 [hep-th]} \BibitemShut {NoStop}%
%%CITATION = ARXIV:1407.7249;%%
\bibitem [{\citenamefont {Lee}\ \emph {et~al.}(2015)\citenamefont {Lee},
  \citenamefont {Lewkowycz}, \citenamefont {Perlmutter},\ and\ \citenamefont
  {Safdi}}]{Lee:2014zaa}%
  \BibitemOpen
  \bibfield  {author} {\bibinfo {author} {\bibfnamefont {Jeongseog}\
  \bibnamefont {Lee}}, \bibinfo {author} {\bibfnamefont {Aitor}\ \bibnamefont
  {Lewkowycz}}, \bibinfo {author} {\bibfnamefont {Eric}\ \bibnamefont
  {Perlmutter}}, \ and\ \bibinfo {author} {\bibfnamefont {Benjamin~R.}\
  \bibnamefont {Safdi}},\ }\bibfield  {title} {\enquote {\bibinfo {title}
  {{Rényi entropy, stationarity, and entanglement of the conformal scalar}},}\
  }\href {\doibase 10.1007/JHEP03(2015)075} {\bibfield  {journal} {\bibinfo
  {journal} {JHEP}\ }\textbf {\bibinfo {volume} {03}},\ \bibinfo {pages} {075}
  (\bibinfo {year} {2015})},\ \Eprint {http://arxiv.org/abs/1407.7816}
  {arXiv:1407.7816 [hep-th]} \BibitemShut {NoStop}%
%%CITATION = ARXIV:1407.7816;%%
\bibitem [{\citenamefont {Lewkowycz}\ and\ \citenamefont
  {Perlmutter}(2015)}]{Lewkowycz:2014jia}%
  \BibitemOpen
  \bibfield  {author} {\bibinfo {author} {\bibfnamefont {Aitor}\ \bibnamefont
  {Lewkowycz}}\ and\ \bibinfo {author} {\bibfnamefont {Eric}\ \bibnamefont
  {Perlmutter}},\ }\bibfield  {title} {\enquote {\bibinfo {title}
  {{Universality in the geometric dependence of Renyi entropy}},}\ }\href
  {\doibase 10.1007/JHEP01(2015)080} {\bibfield  {journal} {\bibinfo  {journal}
  {JHEP}\ }\textbf {\bibinfo {volume} {01}},\ \bibinfo {pages} {080} (\bibinfo
  {year} {2015})},\ \Eprint {http://arxiv.org/abs/1407.8171} {arXiv:1407.8171
  [hep-th]} \BibitemShut {NoStop}%
%%CITATION = ARXIV:1407.8171;%%
\bibitem [{\citenamefont {Bueno}\ \emph
  {et~al.}(2015{\natexlab{a}})\citenamefont {Bueno}, \citenamefont {Myers},\
  and\ \citenamefont {Witczak-Krempa}}]{Bueno:2015rda}%
  \BibitemOpen
  \bibfield  {author} {\bibinfo {author} {\bibfnamefont {Pablo}\ \bibnamefont
  {Bueno}}, \bibinfo {author} {\bibfnamefont {Robert~C.}\ \bibnamefont
  {Myers}}, \ and\ \bibinfo {author} {\bibfnamefont {William}\ \bibnamefont
  {Witczak-Krempa}},\ }\bibfield  {title} {\enquote {\bibinfo {title}
  {{Universality of corner entanglement in conformal field theories}},}\ }\href
  {\doibase 10.1103/PhysRevLett.115.021602} {\bibfield  {journal} {\bibinfo
  {journal} {Phys. Rev. Lett.}\ }\textbf {\bibinfo {volume} {115}},\ \bibinfo
  {pages} {021602} (\bibinfo {year} {2015}{\natexlab{a}})},\ \Eprint
  {http://arxiv.org/abs/1505.04804} {arXiv:1505.04804 [hep-th]} \BibitemShut
  {NoStop}%
%%CITATION = ARXIV:1505.04804;%%
\bibitem [{\citenamefont {Bueno}\ \emph
  {et~al.}(2015{\natexlab{b}})\citenamefont {Bueno}, \citenamefont {Myers},\
  and\ \citenamefont {Witczak-Krempa}}]{Bueno:2015qya}%
  \BibitemOpen
  \bibfield  {author} {\bibinfo {author} {\bibfnamefont {Pablo}\ \bibnamefont
  {Bueno}}, \bibinfo {author} {\bibfnamefont {Robert~C.}\ \bibnamefont
  {Myers}}, \ and\ \bibinfo {author} {\bibfnamefont {William}\ \bibnamefont
  {Witczak-Krempa}},\ }\bibfield  {title} {\enquote {\bibinfo {title}
  {{Universal corner entanglement from twist operators}},}\ }\href {\doibase
  10.1007/JHEP09(2015)091} {\bibfield  {journal} {\bibinfo  {journal} {JHEP}\
  }\textbf {\bibinfo {volume} {09}},\ \bibinfo {pages} {091} (\bibinfo {year}
  {2015}{\natexlab{b}})},\ \Eprint {http://arxiv.org/abs/1507.06997}
  {arXiv:1507.06997 [hep-th]} \BibitemShut {NoStop}%
%%CITATION = ARXIV:1507.06997;%%
\bibitem [{\citenamefont {Bueno}\ and\ \citenamefont
  {Myers}(2015)}]{Bueno:2015lza}%
  \BibitemOpen
  \bibfield  {author} {\bibinfo {author} {\bibfnamefont {Pablo}\ \bibnamefont
  {Bueno}}\ and\ \bibinfo {author} {\bibfnamefont {Robert~C.}\ \bibnamefont
  {Myers}},\ }\bibfield  {title} {\enquote {\bibinfo {title} {{Universal
  entanglement for higher dimensional cones}},}\ }\href {\doibase
  10.1007/JHEP12(2015)168} {\bibfield  {journal} {\bibinfo  {journal} {JHEP}\
  }\textbf {\bibinfo {volume} {12}},\ \bibinfo {pages} {168} (\bibinfo {year}
  {2015})},\ \Eprint {http://arxiv.org/abs/1508.00587} {arXiv:1508.00587
  [hep-th]} \BibitemShut {NoStop}%
%%CITATION = ARXIV:1508.00587;%%
\bibitem [{\citenamefont {Bianchi}\ \emph
  {et~al.}(2016{\natexlab{a}})\citenamefont {Bianchi}, \citenamefont {Meineri},
  \citenamefont {Myers},\ and\ \citenamefont {Smolkin}}]{Bianchi:2015liz}%
  \BibitemOpen
  \bibfield  {author} {\bibinfo {author} {\bibfnamefont {Lorenzo}\ \bibnamefont
  {Bianchi}}, \bibinfo {author} {\bibfnamefont {Marco}\ \bibnamefont
  {Meineri}}, \bibinfo {author} {\bibfnamefont {Robert~C.}\ \bibnamefont
  {Myers}}, \ and\ \bibinfo {author} {\bibfnamefont {Michael}\ \bibnamefont
  {Smolkin}},\ }\bibfield  {title} {\enquote {\bibinfo {title} {{Rényi entropy
  and conformal defects}},}\ }\href {\doibase 10.1007/JHEP07(2016)076}
  {\bibfield  {journal} {\bibinfo  {journal} {JHEP}\ }\textbf {\bibinfo
  {volume} {07}},\ \bibinfo {pages} {076} (\bibinfo {year}
  {2016}{\natexlab{a}})},\ \Eprint {http://arxiv.org/abs/1511.06713}
  {arXiv:1511.06713 [hep-th]} \BibitemShut {NoStop}%
%%CITATION = ARXIV:1511.06713;%%
\bibitem [{\citenamefont {Dong}(2016)}]{Dong:2016wcf}%
  \BibitemOpen
  \bibfield  {author} {\bibinfo {author} {\bibfnamefont {Xi}~\bibnamefont
  {Dong}},\ }\bibfield  {title} {\enquote {\bibinfo {title} {{Shape Dependence
  of Holographic R\'enyi Entropy in Conformal Field Theories}},}\ }\href
  {\doibase 10.1103/PhysRevLett.116.251602} {\bibfield  {journal} {\bibinfo
  {journal} {Phys. Rev. Lett.}\ }\textbf {\bibinfo {volume} {116}},\ \bibinfo
  {pages} {251602} (\bibinfo {year} {2016})},\ \Eprint
  {http://arxiv.org/abs/1602.08493} {arXiv:1602.08493 [hep-th]} \BibitemShut
  {NoStop}%
%%CITATION = ARXIV:1602.08493;%%
\bibitem [{\citenamefont {Bianchi}\ \emph
  {et~al.}(2016{\natexlab{b}})\citenamefont {Bianchi}, \citenamefont {Chapman},
  \citenamefont {Dong}, \citenamefont {Galante}, \citenamefont {Meineri},\ and\
  \citenamefont {Myers}}]{Bianchi:2016xvf}%
  \BibitemOpen
  \bibfield  {author} {\bibinfo {author} {\bibfnamefont {Lorenzo}\ \bibnamefont
  {Bianchi}}, \bibinfo {author} {\bibfnamefont {Shira}\ \bibnamefont
  {Chapman}}, \bibinfo {author} {\bibfnamefont {Xi}~\bibnamefont {Dong}},
  \bibinfo {author} {\bibfnamefont {Damián~A.}\ \bibnamefont {Galante}},
  \bibinfo {author} {\bibfnamefont {Marco}\ \bibnamefont {Meineri}}, \ and\
  \bibinfo {author} {\bibfnamefont {Robert~C.}\ \bibnamefont {Myers}},\
  }\bibfield  {title} {\enquote {\bibinfo {title} {{Shape Dependence of
  Holographic R\'enyi Entropy in General Dimensions}},}\ }\href@noop {} {\
  (\bibinfo {year} {2016}{\natexlab{b}})},\ \Eprint
  {http://arxiv.org/abs/1607.07418} {arXiv:1607.07418 [hep-th]} \BibitemShut
  {NoStop}%
%%CITATION = ARXIV:1607.07418;%%
\bibitem [{\citenamefont {Headrick}(2010)}]{Headrick:2010zt}%
  \BibitemOpen
  \bibfield  {author} {\bibinfo {author} {\bibfnamefont {Matthew}\ \bibnamefont
  {Headrick}},\ }\bibfield  {title} {\enquote {\bibinfo {title} {{Entanglement
  Renyi entropies in holographic theories}},}\ }\href {\doibase
  10.1103/PhysRevD.82.126010} {\bibfield  {journal} {\bibinfo  {journal}
  {Phys.Rev.}\ }\textbf {\bibinfo {volume} {D82}},\ \bibinfo {pages} {126010}
  (\bibinfo {year} {2010})},\ \Eprint {http://arxiv.org/abs/1006.0047}
  {arXiv:1006.0047 [hep-th]} \BibitemShut {NoStop}%
%%CITATION = ARXIV:1006.0047;%%
\bibitem [{\citenamefont {Hung}\ \emph {et~al.}(2011)\citenamefont {Hung},
  \citenamefont {Myers}, \citenamefont {Smolkin},\ and\ \citenamefont
  {Yale}}]{Hung:2011nu}%
  \BibitemOpen
  \bibfield  {author} {\bibinfo {author} {\bibfnamefont {Ling-Yan}\
  \bibnamefont {Hung}}, \bibinfo {author} {\bibfnamefont {Robert~C.}\
  \bibnamefont {Myers}}, \bibinfo {author} {\bibfnamefont {Michael}\
  \bibnamefont {Smolkin}}, \ and\ \bibinfo {author} {\bibfnamefont {Alexandre}\
  \bibnamefont {Yale}},\ }\bibfield  {title} {\enquote {\bibinfo {title}
  {{Holographic Calculations of Renyi Entropy}},}\ }\href {\doibase
  10.1007/JHEP12(2011)047} {\bibfield  {journal} {\bibinfo  {journal} {JHEP}\
  }\textbf {\bibinfo {volume} {12}},\ \bibinfo {pages} {047} (\bibinfo {year}
  {2011})},\ \Eprint {http://arxiv.org/abs/1110.1084} {arXiv:1110.1084
  [hep-th]} \BibitemShut {NoStop}%
%%CITATION = ARXIV:1110.1084;%%
\bibitem [{\citenamefont {Fursaev}(2012)}]{Fursaev:2012mp}%
  \BibitemOpen
  \bibfield  {author} {\bibinfo {author} {\bibfnamefont {D.~V.}\ \bibnamefont
  {Fursaev}},\ }\bibfield  {title} {\enquote {\bibinfo {title} {{Entanglement
  Renyi Entropies in Conformal Field Theories and Holography}},}\ }\href
  {\doibase 10.1007/JHEP05(2012)080} {\bibfield  {journal} {\bibinfo  {journal}
  {JHEP}\ }\textbf {\bibinfo {volume} {05}},\ \bibinfo {pages} {080} (\bibinfo
  {year} {2012})},\ \Eprint {http://arxiv.org/abs/1201.1702} {arXiv:1201.1702
  [hep-th]} \BibitemShut {NoStop}%
%%CITATION = ARXIV:1201.1702;%%
\bibitem [{\citenamefont {Faulkner}(2013)}]{Faulkner:2013yia}%
  \BibitemOpen
  \bibfield  {author} {\bibinfo {author} {\bibfnamefont {Thomas}\ \bibnamefont
  {Faulkner}},\ }\bibfield  {title} {\enquote {\bibinfo {title} {{The
  Entanglement Renyi Entropies of Disjoint Intervals in AdS/CFT}},}\
  }\href@noop {} {\  (\bibinfo {year} {2013})},\ \Eprint
  {http://arxiv.org/abs/1303.7221} {arXiv:1303.7221 [hep-th]} \BibitemShut
  {NoStop}%
%%CITATION = ARXIV:1303.7221;%%
\bibitem [{\citenamefont {Galante}\ and\ \citenamefont
  {Myers}(2013)}]{Galante:2013wta}%
  \BibitemOpen
  \bibfield  {author} {\bibinfo {author} {\bibfnamefont {Damián~A.}\
  \bibnamefont {Galante}}\ and\ \bibinfo {author} {\bibfnamefont {Robert~C.}\
  \bibnamefont {Myers}},\ }\bibfield  {title} {\enquote {\bibinfo {title}
  {{Holographic Renyi entropies at finite coupling}},}\ }\href {\doibase
  10.1007/JHEP08(2013)063} {\bibfield  {journal} {\bibinfo  {journal} {JHEP}\
  }\textbf {\bibinfo {volume} {08}},\ \bibinfo {pages} {063} (\bibinfo {year}
  {2013})},\ \Eprint {http://arxiv.org/abs/1305.7191} {arXiv:1305.7191
  [hep-th]} \BibitemShut {NoStop}%
%%CITATION = ARXIV:1305.7191;%%
\bibitem [{\citenamefont {Belin}\ \emph
  {et~al.}(2013{\natexlab{a}})\citenamefont {Belin}, \citenamefont {Maloney},\
  and\ \citenamefont {Matsuura}}]{Belin:2013dva}%
  \BibitemOpen
  \bibfield  {author} {\bibinfo {author} {\bibfnamefont {Alexandre}\
  \bibnamefont {Belin}}, \bibinfo {author} {\bibfnamefont {Alexander}\
  \bibnamefont {Maloney}}, \ and\ \bibinfo {author} {\bibfnamefont {Shunji}\
  \bibnamefont {Matsuura}},\ }\bibfield  {title} {\enquote {\bibinfo {title}
  {{Holographic Phases of Renyi Entropies}},}\ }\href {\doibase
  10.1007/JHEP12(2013)050} {\bibfield  {journal} {\bibinfo  {journal} {JHEP}\
  }\textbf {\bibinfo {volume} {12}},\ \bibinfo {pages} {050} (\bibinfo {year}
  {2013}{\natexlab{a}})},\ \Eprint {http://arxiv.org/abs/1306.2640}
  {arXiv:1306.2640 [hep-th]} \BibitemShut {NoStop}%
%%CITATION = ARXIV:1306.2640;%%
\bibitem [{\citenamefont {Barrella}\ \emph {et~al.}(2013)\citenamefont
  {Barrella}, \citenamefont {Dong}, \citenamefont {Hartnoll},\ and\
  \citenamefont {Martin}}]{Barrella:2013wja}%
  \BibitemOpen
  \bibfield  {author} {\bibinfo {author} {\bibfnamefont {Taylor}\ \bibnamefont
  {Barrella}}, \bibinfo {author} {\bibfnamefont {Xi}~\bibnamefont {Dong}},
  \bibinfo {author} {\bibfnamefont {Sean~A.}\ \bibnamefont {Hartnoll}}, \ and\
  \bibinfo {author} {\bibfnamefont {Victoria~L.}\ \bibnamefont {Martin}},\
  }\bibfield  {title} {\enquote {\bibinfo {title} {{Holographic entanglement
  beyond classical gravity}},}\ }\href {\doibase 10.1007/JHEP09(2013)109}
  {\bibfield  {journal} {\bibinfo  {journal} {JHEP}\ }\textbf {\bibinfo
  {volume} {09}},\ \bibinfo {pages} {109} (\bibinfo {year} {2013})},\ \Eprint
  {http://arxiv.org/abs/1306.4682} {arXiv:1306.4682 [hep-th]} \BibitemShut
  {NoStop}%
%%CITATION = ARXIV:1306.4682;%%
\bibitem [{\citenamefont {Chen}\ \emph {et~al.}(2014)\citenamefont {Chen},
  \citenamefont {Long},\ and\ \citenamefont {Zhang}}]{Chen:2013dxa}%
  \BibitemOpen
  \bibfield  {author} {\bibinfo {author} {\bibfnamefont {Bin}\ \bibnamefont
  {Chen}}, \bibinfo {author} {\bibfnamefont {Jiang}\ \bibnamefont {Long}}, \
  and\ \bibinfo {author} {\bibfnamefont {Jia-ju}\ \bibnamefont {Zhang}},\
  }\bibfield  {title} {\enquote {\bibinfo {title} {{Holographic Rényi entropy
  for CFT with W symmetry}},}\ }\href {\doibase 10.1007/JHEP04(2014)041}
  {\bibfield  {journal} {\bibinfo  {journal} {JHEP}\ }\textbf {\bibinfo
  {volume} {04}},\ \bibinfo {pages} {041} (\bibinfo {year} {2014})},\ \Eprint
  {http://arxiv.org/abs/1312.5510} {arXiv:1312.5510 [hep-th]} \BibitemShut
  {NoStop}%
%%CITATION = ARXIV:1312.5510;%%
\bibitem [{\citenamefont {Belin}\ \emph
  {et~al.}(2013{\natexlab{b}})\citenamefont {Belin}, \citenamefont {Hung},
  \citenamefont {Maloney}, \citenamefont {Matsuura}, \citenamefont {Myers},\
  and\ \citenamefont {Sierens}}]{Belin:2013uta}%
  \BibitemOpen
  \bibfield  {author} {\bibinfo {author} {\bibfnamefont {Alexandre}\
  \bibnamefont {Belin}}, \bibinfo {author} {\bibfnamefont {Ling-Yan}\
  \bibnamefont {Hung}}, \bibinfo {author} {\bibfnamefont {Alexander}\
  \bibnamefont {Maloney}}, \bibinfo {author} {\bibfnamefont {Shunji}\
  \bibnamefont {Matsuura}}, \bibinfo {author} {\bibfnamefont {Robert~C.}\
  \bibnamefont {Myers}}, \ and\ \bibinfo {author} {\bibfnamefont {Todd}\
  \bibnamefont {Sierens}},\ }\bibfield  {title} {\enquote {\bibinfo {title}
  {{Holographic Charged Renyi Entropies}},}\ }\href {\doibase
  10.1007/JHEP12(2013)059} {\bibfield  {journal} {\bibinfo  {journal} {JHEP}\
  }\textbf {\bibinfo {volume} {12}},\ \bibinfo {pages} {059} (\bibinfo {year}
  {2013}{\natexlab{b}})},\ \Eprint {http://arxiv.org/abs/1310.4180}
  {arXiv:1310.4180 [hep-th]} \BibitemShut {NoStop}%
%%CITATION = ARXIV:1310.4180;%%
\bibitem [{\citenamefont {Belin}\ \emph {et~al.}(2015)\citenamefont {Belin},
  \citenamefont {Hung}, \citenamefont {Maloney},\ and\ \citenamefont
  {Matsuura}}]{Belin:2014mva}%
  \BibitemOpen
  \bibfield  {author} {\bibinfo {author} {\bibfnamefont {Alexandre}\
  \bibnamefont {Belin}}, \bibinfo {author} {\bibfnamefont {Ling-Yan}\
  \bibnamefont {Hung}}, \bibinfo {author} {\bibfnamefont {Alexander}\
  \bibnamefont {Maloney}}, \ and\ \bibinfo {author} {\bibfnamefont {Shunji}\
  \bibnamefont {Matsuura}},\ }\bibfield  {title} {\enquote {\bibinfo {title}
  {{Charged Renyi entropies and holographic superconductors}},}\ }\href
  {\doibase 10.1007/JHEP01(2015)059} {\bibfield  {journal} {\bibinfo  {journal}
  {JHEP}\ }\textbf {\bibinfo {volume} {01}},\ \bibinfo {pages} {059} (\bibinfo
  {year} {2015})},\ \Eprint {http://arxiv.org/abs/1407.5630} {arXiv:1407.5630
  [hep-th]} \BibitemShut {NoStop}%
%%CITATION = ARXIV:1407.5630;%%
\bibitem [{\citenamefont {Nishioka}\ and\ \citenamefont
  {Yaakov}(2013)}]{Nishioka:2013haa}%
  \BibitemOpen
  \bibfield  {author} {\bibinfo {author} {\bibfnamefont {Tatsuma}\ \bibnamefont
  {Nishioka}}\ and\ \bibinfo {author} {\bibfnamefont {Itamar}\ \bibnamefont
  {Yaakov}},\ }\bibfield  {title} {\enquote {\bibinfo {title} {{Supersymmetric
  Renyi Entropy}},}\ }\href {\doibase 10.1007/JHEP10(2013)155} {\bibfield
  {journal} {\bibinfo  {journal} {JHEP}\ }\textbf {\bibinfo {volume} {10}},\
  \bibinfo {pages} {155} (\bibinfo {year} {2013})},\ \Eprint
  {http://arxiv.org/abs/1306.2958} {arXiv:1306.2958 [hep-th]} \BibitemShut
  {NoStop}%
%%CITATION = ARXIV:1306.2958;%%
\bibitem [{\citenamefont {Alday}\ \emph {et~al.}(2015)\citenamefont {Alday},
  \citenamefont {Richmond},\ and\ \citenamefont {Sparks}}]{Alday:2014fsa}%
  \BibitemOpen
  \bibfield  {author} {\bibinfo {author} {\bibfnamefont {Luis~F.}\ \bibnamefont
  {Alday}}, \bibinfo {author} {\bibfnamefont {Paul}\ \bibnamefont {Richmond}},
  \ and\ \bibinfo {author} {\bibfnamefont {James}\ \bibnamefont {Sparks}},\
  }\bibfield  {title} {\enquote {\bibinfo {title} {{The holographic
  supersymmetric Renyi entropy in five dimensions}},}\ }\href {\doibase
  10.1007/JHEP02(2015)102} {\bibfield  {journal} {\bibinfo  {journal} {JHEP}\
  }\textbf {\bibinfo {volume} {02}},\ \bibinfo {pages} {102} (\bibinfo {year}
  {2015})},\ \Eprint {http://arxiv.org/abs/1410.0899} {arXiv:1410.0899
  [hep-th]} \BibitemShut {NoStop}%
%%CITATION = ARXIV:1410.0899;%%
\bibitem [{\citenamefont {Giveon}\ and\ \citenamefont
  {Kutasov}(2016)}]{Giveon:2015cgs}%
  \BibitemOpen
  \bibfield  {author} {\bibinfo {author} {\bibfnamefont {Amit}\ \bibnamefont
  {Giveon}}\ and\ \bibinfo {author} {\bibfnamefont {David}\ \bibnamefont
  {Kutasov}},\ }\bibfield  {title} {\enquote {\bibinfo {title} {{Supersymmetric
  Renyi entropy in CFT$_{2}$ and AdS$_{3}$}},}\ }\href {\doibase
  10.1007/JHEP01(2016)042} {\bibfield  {journal} {\bibinfo  {journal} {JHEP}\
  }\textbf {\bibinfo {volume} {01}},\ \bibinfo {pages} {042} (\bibinfo {year}
  {2016})},\ \Eprint {http://arxiv.org/abs/1510.08872} {arXiv:1510.08872
  [hep-th]} \BibitemShut {NoStop}%
%%CITATION = ARXIV:1510.08872;%%
\bibitem [{\citenamefont {Bombelli}\ \emph {et~al.}(1986)\citenamefont
  {Bombelli}, \citenamefont {Koul}, \citenamefont {Lee},\ and\ \citenamefont
  {Sorkin}}]{Bombelli:1986rw}%
  \BibitemOpen
  \bibfield  {author} {\bibinfo {author} {\bibfnamefont {Luca}\ \bibnamefont
  {Bombelli}}, \bibinfo {author} {\bibfnamefont {Rabinder~K.}\ \bibnamefont
  {Koul}}, \bibinfo {author} {\bibfnamefont {Joohan}\ \bibnamefont {Lee}}, \
  and\ \bibinfo {author} {\bibfnamefont {Rafael~D.}\ \bibnamefont {Sorkin}},\
  }\bibfield  {title} {\enquote {\bibinfo {title} {{A Quantum Source of Entropy
  for Black Holes}},}\ }\href {\doibase 10.1103/PhysRevD.34.373} {\bibfield
  {journal} {\bibinfo  {journal} {Phys. Rev.}\ }\textbf {\bibinfo {volume}
  {D34}},\ \bibinfo {pages} {373--383} (\bibinfo {year} {1986})}\BibitemShut
  {NoStop}%
%%CITATION = PHRVA,D34,373;%%
\bibitem [{\citenamefont {Srednicki}(1993)}]{Srednicki:1993im}%
  \BibitemOpen
  \bibfield  {author} {\bibinfo {author} {\bibfnamefont {Mark}\ \bibnamefont
  {Srednicki}},\ }\bibfield  {title} {\enquote {\bibinfo {title} {{Entropy and
  area}},}\ }\href {\doibase 10.1103/PhysRevLett.71.666} {\bibfield  {journal}
  {\bibinfo  {journal} {Phys. Rev. Lett.}\ }\textbf {\bibinfo {volume} {71}},\
  \bibinfo {pages} {666--669} (\bibinfo {year} {1993})},\ \Eprint
  {http://arxiv.org/abs/hep-th/9303048} {arXiv:hep-th/9303048 [hep-th]}
  \BibitemShut {NoStop}%
%%CITATION = HEP-TH/9303048;%%
\bibitem [{\citenamefont {Hastings}(2007{\natexlab{a}})}]{PhysRevB.76.035114}%
  \BibitemOpen
  \bibfield  {author} {\bibinfo {author} {\bibfnamefont {Matthew~B.}\
  \bibnamefont {Hastings}},\ }\bibfield  {title} {\enquote {\bibinfo {title}
  {Entropy and entanglement in quantum ground states},}\ }\href {\doibase
  10.1103/PhysRevB.76.035114} {\bibfield  {journal} {\bibinfo  {journal} {Phys.
  Rev. B}\ }\textbf {\bibinfo {volume} {76}},\ \bibinfo {pages} {035114}
  (\bibinfo {year} {2007}{\natexlab{a}})},\ \Eprint
  {http://arxiv.org/abs/cond-mat/0701055} {arXiv:cond-mat/0701055
  [cond-mat.str-el]} \BibitemShut {NoStop}%
\bibitem [{\citenamefont {Wolf}\ \emph {et~al.}(2008)\citenamefont {Wolf},
  \citenamefont {Verstraete}, \citenamefont {Hastings},\ and\ \citenamefont
  {Cirac}}]{PhysRevLett.100.070502}%
  \BibitemOpen
  \bibfield  {author} {\bibinfo {author} {\bibfnamefont {Michael~M.}\
  \bibnamefont {Wolf}}, \bibinfo {author} {\bibfnamefont {Frank}\ \bibnamefont
  {Verstraete}}, \bibinfo {author} {\bibfnamefont {Matthew~B.}\ \bibnamefont
  {Hastings}}, \ and\ \bibinfo {author} {\bibfnamefont {J.~Ignacio}\
  \bibnamefont {Cirac}},\ }\bibfield  {title} {\enquote {\bibinfo {title} {Area
  laws in quantum systems: Mutual information and correlations},}\ }\href
  {\doibase 10.1103/PhysRevLett.100.070502} {\bibfield  {journal} {\bibinfo
  {journal} {Phys. Rev. Lett.}\ }\textbf {\bibinfo {volume} {100}},\ \bibinfo
  {pages} {070502} (\bibinfo {year} {2008})},\ \Eprint
  {http://arxiv.org/abs/0704.3906} {arXiv:0704.3906 [quant-ph]} \BibitemShut
  {NoStop}%
\bibitem [{\citenamefont
  {Hastings}(2007{\natexlab{b}})}]{1742-5468-2007-08-P08024}%
  \BibitemOpen
  \bibfield  {author} {\bibinfo {author} {\bibfnamefont {Matthew~B.}\
  \bibnamefont {Hastings}},\ }\bibfield  {title} {\enquote {\bibinfo {title}
  {An area law for one-dimensional quantum systems},}\ }\href
  {http://stacks.iop.org/1742-5468/2007/i=08/a=P08024} {\bibfield  {journal}
  {\bibinfo  {journal} {Journal of Statistical Mechanics: Theory and
  Experiment}\ }\textbf {\bibinfo {volume} {2007}},\ \bibinfo {pages} {P08024}
  (\bibinfo {year} {2007}{\natexlab{b}})},\ \Eprint
  {http://arxiv.org/abs/0705.2024} {arXiv:0705.2024 [quant-ph]} \BibitemShut
  {NoStop}%
\bibitem [{\citenamefont {Eisert}\ \emph {et~al.}(2010)\citenamefont {Eisert},
  \citenamefont {Cramer},\ and\ \citenamefont {Plenio}}]{Eisert:2008ur}%
  \BibitemOpen
  \bibfield  {author} {\bibinfo {author} {\bibfnamefont {J.}~\bibnamefont
  {Eisert}}, \bibinfo {author} {\bibfnamefont {M.}~\bibnamefont {Cramer}}, \
  and\ \bibinfo {author} {\bibfnamefont {M.~B.}\ \bibnamefont {Plenio}},\
  }\bibfield  {title} {\enquote {\bibinfo {title} {{Area laws for the
  entanglement entropy - a review}},}\ }\href {\doibase
  10.1103/RevModPhys.82.277} {\bibfield  {journal} {\bibinfo  {journal} {Rev.
  Mod. Phys.}\ }\textbf {\bibinfo {volume} {82}},\ \bibinfo {pages} {277--306}
  (\bibinfo {year} {2010})},\ \Eprint {http://arxiv.org/abs/0808.3773}
  {arXiv:0808.3773 [quant-ph]} \BibitemShut {NoStop}%
%%CITATION = ARXIV:0808.3773;%%
\bibitem [{\citenamefont {Vilenkin}(1981)}]{Vilenkin:1981zs}%
  \BibitemOpen
  \bibfield  {author} {\bibinfo {author} {\bibfnamefont {A.}~\bibnamefont
  {Vilenkin}},\ }\bibfield  {title} {\enquote {\bibinfo {title} {{Gravitational
  Field of Vacuum Domain Walls and Strings}},}\ }\href {\doibase
  10.1103/PhysRevD.23.852} {\bibfield  {journal} {\bibinfo  {journal} {Phys.
  Rev.}\ }\textbf {\bibinfo {volume} {D23}},\ \bibinfo {pages} {852--857}
  (\bibinfo {year} {1981})}\BibitemShut {NoStop}%
%%CITATION = PHRVA,D23,852;%%
\bibitem [{\citenamefont {Beck}\ and\ \citenamefont
  {Schögl}(1993)}]{10.1017/CBO9780511524585}%
  \BibitemOpen
  \bibfield  {author} {\bibinfo {author} {\bibfnamefont {Christian}\
  \bibnamefont {Beck}}\ and\ \bibinfo {author} {\bibfnamefont {Friedrich}\
  \bibnamefont {Schögl}},\ }\href {\doibase 10.1017/CBO9780511524585} {\emph
  {\bibinfo {title} {Thermodynamics of Chaotic Systems}}}\ (\bibinfo
  {publisher} {Cambridge University Press},\ \bibinfo {year}
  {1993})\BibitemShut {NoStop}%
\bibitem [{\citenamefont {Baez}(2011)}]{Baez:2011}%
  \BibitemOpen
  \bibfield  {author} {\bibinfo {author} {\bibfnamefont {John~C.}\ \bibnamefont
  {Baez}},\ }\bibfield  {title} {\enquote {\bibinfo {title} {{Rényi Entropy
  and Free Energy}},}\ }\href@noop {} {\  (\bibinfo {year} {2011})},\ \Eprint
  {http://arxiv.org/abs/1102.2098} {arXiv:1102.2098 [quant-ph]} \BibitemShut
  {NoStop}%
\bibitem [{\citenamefont {Jafferis}\ and\ \citenamefont
  {Suh}(2014)}]{Jafferis:2014lza}%
  \BibitemOpen
  \bibfield  {author} {\bibinfo {author} {\bibfnamefont {Daniel~L.}\
  \bibnamefont {Jafferis}}\ and\ \bibinfo {author} {\bibfnamefont
  {S.~Josephine}\ \bibnamefont {Suh}},\ }\bibfield  {title} {\enquote {\bibinfo
  {title} {{The Gravity Duals of Modular Hamiltonians}},}\ }\href@noop {} {\
  (\bibinfo {year} {2014})},\ \Eprint {http://arxiv.org/abs/1412.8465}
  {arXiv:1412.8465 [hep-th]} \BibitemShut {NoStop}%
%%CITATION = ARXIV:1412.8465;%%
\bibitem [{\citenamefont {Jafferis}\ \emph {et~al.}(2016)\citenamefont
  {Jafferis}, \citenamefont {Lewkowycz}, \citenamefont {Maldacena},\ and\
  \citenamefont {Suh}}]{Jafferis:2015del}%
  \BibitemOpen
  \bibfield  {author} {\bibinfo {author} {\bibfnamefont {Daniel~L.}\
  \bibnamefont {Jafferis}}, \bibinfo {author} {\bibfnamefont {Aitor}\
  \bibnamefont {Lewkowycz}}, \bibinfo {author} {\bibfnamefont {Juan}\
  \bibnamefont {Maldacena}}, \ and\ \bibinfo {author} {\bibfnamefont
  {S.~Josephine}\ \bibnamefont {Suh}},\ }\bibfield  {title} {\enquote {\bibinfo
  {title} {{Relative entropy equals bulk relative entropy}},}\ }\href {\doibase
  10.1007/JHEP06(2016)004} {\bibfield  {journal} {\bibinfo  {journal} {JHEP}\
  }\textbf {\bibinfo {volume} {06}},\ \bibinfo {pages} {004} (\bibinfo {year}
  {2016})},\ \Eprint {http://arxiv.org/abs/1512.06431} {arXiv:1512.06431
  [hep-th]} \BibitemShut {NoStop}%
%%CITATION = ARXIV:1512.06431;%%
\bibitem [{Note1()}]{Note1}%
  \BibitemOpen
  \bibinfo {note} {We thank E.~Perlmutter for reminding us of this entropy
  inequality.}\BibitemShut {Stop}%
\bibitem [{\citenamefont {Wald}(1993)}]{Wald:1993nt}%
  \BibitemOpen
  \bibfield  {author} {\bibinfo {author} {\bibfnamefont {Robert~M.}\
  \bibnamefont {Wald}},\ }\bibfield  {title} {\enquote {\bibinfo {title}
  {{Black hole entropy is the Noether charge}},}\ }\href {\doibase
  10.1103/PhysRevD.48.R3427} {\bibfield  {journal} {\bibinfo  {journal}
  {Phys.Rev.}\ }\textbf {\bibinfo {volume} {D48}},\ \bibinfo {pages}
  {3427--3431} (\bibinfo {year} {1993})},\ \Eprint
  {http://arxiv.org/abs/gr-qc/9307038} {arXiv:gr-qc/9307038 [gr-qc]}
  \BibitemShut {NoStop}%
%%CITATION = GR-QC/9307038;%%
\bibitem [{\citenamefont {Jacobson}\ \emph {et~al.}(1994)\citenamefont
  {Jacobson}, \citenamefont {Kang},\ and\ \citenamefont
  {Myers}}]{Jacobson:1993vj}%
  \BibitemOpen
  \bibfield  {author} {\bibinfo {author} {\bibfnamefont {Ted}\ \bibnamefont
  {Jacobson}}, \bibinfo {author} {\bibfnamefont {Gungwon}\ \bibnamefont
  {Kang}}, \ and\ \bibinfo {author} {\bibfnamefont {Robert~C.}\ \bibnamefont
  {Myers}},\ }\bibfield  {title} {\enquote {\bibinfo {title} {{On black hole
  entropy}},}\ }\href {\doibase 10.1103/PhysRevD.49.6587} {\bibfield  {journal}
  {\bibinfo  {journal} {Phys.Rev.}\ }\textbf {\bibinfo {volume} {D49}},\
  \bibinfo {pages} {6587--6598} (\bibinfo {year} {1994})},\ \Eprint
  {http://arxiv.org/abs/gr-qc/9312023} {arXiv:gr-qc/9312023 [gr-qc]}
  \BibitemShut {NoStop}%
%%CITATION = GR-QC/9312023;%%
\bibitem [{\citenamefont {Iyer}\ and\ \citenamefont
  {Wald}(1994)}]{Iyer:1994ys}%
  \BibitemOpen
  \bibfield  {author} {\bibinfo {author} {\bibfnamefont {Vivek}\ \bibnamefont
  {Iyer}}\ and\ \bibinfo {author} {\bibfnamefont {Robert~M.}\ \bibnamefont
  {Wald}},\ }\bibfield  {title} {\enquote {\bibinfo {title} {{Some properties
  of Noether charge and a proposal for dynamical black hole entropy}},}\ }\href
  {\doibase 10.1103/PhysRevD.50.846} {\bibfield  {journal} {\bibinfo  {journal}
  {Phys.Rev.}\ }\textbf {\bibinfo {volume} {D50}},\ \bibinfo {pages} {846--864}
  (\bibinfo {year} {1994})},\ \Eprint {http://arxiv.org/abs/gr-qc/9403028}
  {arXiv:gr-qc/9403028 [gr-qc]} \BibitemShut {NoStop}%
%%CITATION = GR-QC/9403028;%%
\bibitem [{\citenamefont {Engelhardt}\ and\ \citenamefont
  {Wall}(2015)}]{Engelhardt:2014gca}%
  \BibitemOpen
  \bibfield  {author} {\bibinfo {author} {\bibfnamefont {Netta}\ \bibnamefont
  {Engelhardt}}\ and\ \bibinfo {author} {\bibfnamefont {Aron~C.}\ \bibnamefont
  {Wall}},\ }\bibfield  {title} {\enquote {\bibinfo {title} {{Quantum Extremal
  Surfaces: Holographic Entanglement Entropy beyond the Classical Regime}},}\
  }\href {\doibase 10.1007/JHEP01(2015)073} {\bibfield  {journal} {\bibinfo
  {journal} {JHEP}\ }\textbf {\bibinfo {volume} {01}},\ \bibinfo {pages} {073}
  (\bibinfo {year} {2015})},\ \Eprint {http://arxiv.org/abs/1408.3203}
  {arXiv:1408.3203 [hep-th]} \BibitemShut {NoStop}%
%%CITATION = ARXIV:1408.3203;%%
\bibitem [{\citenamefont {Haehl}\ \emph {et~al.}(2015)\citenamefont {Haehl},
  \citenamefont {Hartman}, \citenamefont {Marolf}, \citenamefont {Maxfield},\
  and\ \citenamefont {Rangamani}}]{Haehl:2014zoa}%
  \BibitemOpen
  \bibfield  {author} {\bibinfo {author} {\bibfnamefont {Felix~M.}\
  \bibnamefont {Haehl}}, \bibinfo {author} {\bibfnamefont {Thomas}\
  \bibnamefont {Hartman}}, \bibinfo {author} {\bibfnamefont {Donald}\
  \bibnamefont {Marolf}}, \bibinfo {author} {\bibfnamefont {Henry}\
  \bibnamefont {Maxfield}}, \ and\ \bibinfo {author} {\bibfnamefont {Mukund}\
  \bibnamefont {Rangamani}},\ }\bibfield  {title} {\enquote {\bibinfo {title}
  {{Topological aspects of generalized gravitational entropy}},}\ }\href
  {\doibase 10.1007/JHEP05(2015)023} {\bibfield  {journal} {\bibinfo  {journal}
  {JHEP}\ }\textbf {\bibinfo {volume} {05}},\ \bibinfo {pages} {023} (\bibinfo
  {year} {2015})},\ \Eprint {http://arxiv.org/abs/1412.7561} {arXiv:1412.7561
  [hep-th]} \BibitemShut {NoStop}%
%%CITATION = ARXIV:1412.7561;%%
\bibitem [{\citenamefont {Osborn}\ and\ \citenamefont
  {Petkou}(1994)}]{Osborn:1993cr}%
  \BibitemOpen
  \bibfield  {author} {\bibinfo {author} {\bibfnamefont {H.}~\bibnamefont
  {Osborn}}\ and\ \bibinfo {author} {\bibfnamefont {A.~C.}\ \bibnamefont
  {Petkou}},\ }\bibfield  {title} {\enquote {\bibinfo {title} {{Implications of
  conformal invariance in field theories for general dimensions}},}\ }\href
  {\doibase 10.1006/aphy.1994.1045} {\bibfield  {journal} {\bibinfo  {journal}
  {Annals Phys.}\ }\textbf {\bibinfo {volume} {231}},\ \bibinfo {pages}
  {311--362} (\bibinfo {year} {1994})},\ \Eprint
  {http://arxiv.org/abs/hep-th/9307010} {arXiv:hep-th/9307010 [hep-th]}
  \BibitemShut {NoStop}%
%%CITATION = HEP-TH/9307010;%%
\bibitem [{\citenamefont {Erdmenger}\ and\ \citenamefont
  {Osborn}(1997)}]{Erdmenger:1996yc}%
  \BibitemOpen
  \bibfield  {author} {\bibinfo {author} {\bibfnamefont {J.}~\bibnamefont
  {Erdmenger}}\ and\ \bibinfo {author} {\bibfnamefont {H.}~\bibnamefont
  {Osborn}},\ }\bibfield  {title} {\enquote {\bibinfo {title} {{Conserved
  currents and the energy momentum tensor in conformally invariant theories for
  general dimensions}},}\ }\href {\doibase 10.1016/S0550-3213(96)00545-7}
  {\bibfield  {journal} {\bibinfo  {journal} {Nucl. Phys.}\ }\textbf {\bibinfo
  {volume} {B483}},\ \bibinfo {pages} {431--474} (\bibinfo {year} {1997})},\
  \Eprint {http://arxiv.org/abs/hep-th/9605009} {arXiv:hep-th/9605009 [hep-th]}
  \BibitemShut {NoStop}%
%%CITATION = HEP-TH/9605009;%%
\bibitem [{\citenamefont {'t~Hooft}(1993)}]{'tHooft:1993gx}%
  \BibitemOpen
  \bibfield  {author} {\bibinfo {author} {\bibfnamefont {Gerard}\ \bibnamefont
  {'t~Hooft}},\ }\bibfield  {title} {\enquote {\bibinfo {title} {{Dimensional
  reduction in quantum gravity}},}\ }in\ \href@noop {} {\emph {\bibinfo
  {booktitle} {{Salamfest}}}}\ (\bibinfo {year} {1993})\ pp.\ \bibinfo {pages}
  {0284--296},\ \Eprint {http://arxiv.org/abs/gr-qc/9310026}
  {arXiv:gr-qc/9310026 [gr-qc]} \BibitemShut {NoStop}%
%%CITATION = GR-QC/9310026;%%
\bibitem [{\citenamefont {Susskind}(1995)}]{Susskind:1994vu}%
  \BibitemOpen
  \bibfield  {author} {\bibinfo {author} {\bibfnamefont {Leonard}\ \bibnamefont
  {Susskind}},\ }\bibfield  {title} {\enquote {\bibinfo {title} {{The World as
  a hologram}},}\ }\href {\doibase 10.1063/1.531249} {\bibfield  {journal}
  {\bibinfo  {journal} {J. Math. Phys.}\ }\textbf {\bibinfo {volume} {36}},\
  \bibinfo {pages} {6377--6396} (\bibinfo {year} {1995})},\ \Eprint
  {http://arxiv.org/abs/hep-th/9409089} {arXiv:hep-th/9409089 [hep-th]}
  \BibitemShut {NoStop}%
%%CITATION = HEP-TH/9409089;%%
\bibitem [{\citenamefont {Jacobson}(1995)}]{Jacobson:1995ab}%
  \BibitemOpen
  \bibfield  {author} {\bibinfo {author} {\bibfnamefont {Ted}\ \bibnamefont
  {Jacobson}},\ }\bibfield  {title} {\enquote {\bibinfo {title}
  {{Thermodynamics of space-time: The Einstein equation of state}},}\ }\href
  {\doibase 10.1103/PhysRevLett.75.1260} {\bibfield  {journal} {\bibinfo
  {journal} {Phys. Rev. Lett.}\ }\textbf {\bibinfo {volume} {75}},\ \bibinfo
  {pages} {1260--1263} (\bibinfo {year} {1995})},\ \Eprint
  {http://arxiv.org/abs/gr-qc/9504004} {arXiv:gr-qc/9504004 [gr-qc]}
  \BibitemShut {NoStop}%
%%CITATION = GR-QC/9504004;%%
\bibitem [{\citenamefont {Van~Raamsdonk}(2010)}]{VanRaamsdonk:2010pw}%
  \BibitemOpen
  \bibfield  {author} {\bibinfo {author} {\bibfnamefont {Mark}\ \bibnamefont
  {Van~Raamsdonk}},\ }\bibfield  {title} {\enquote {\bibinfo {title} {{Building
  up spacetime with quantum entanglement}},}\ }\href {\doibase
  10.1007/s10714-010-1034-0, 10.1142/S0218271810018529} {\bibfield  {journal}
  {\bibinfo  {journal} {Gen.Rel.Grav.}\ }\textbf {\bibinfo {volume} {42}},\
  \bibinfo {pages} {2323--2329} (\bibinfo {year} {2010})},\ \Eprint
  {http://arxiv.org/abs/1005.3035} {arXiv:1005.3035 [hep-th]} \BibitemShut
  {NoStop}%
%%CITATION = ARXIV:1005.3035;%%
\bibitem [{\citenamefont {Maldacena}\ and\ \citenamefont
  {Susskind}(2013)}]{Maldacena:2013xja}%
  \BibitemOpen
  \bibfield  {author} {\bibinfo {author} {\bibfnamefont {Juan}\ \bibnamefont
  {Maldacena}}\ and\ \bibinfo {author} {\bibfnamefont {Leonard}\ \bibnamefont
  {Susskind}},\ }\bibfield  {title} {\enquote {\bibinfo {title} {{Cool horizons
  for entangled black holes}},}\ }\href {\doibase 10.1002/prop.201300020}
  {\bibfield  {journal} {\bibinfo  {journal} {Fortsch. Phys.}\ }\textbf
  {\bibinfo {volume} {61}},\ \bibinfo {pages} {781--811} (\bibinfo {year}
  {2013})},\ \Eprint {http://arxiv.org/abs/1306.0533} {arXiv:1306.0533
  [hep-th]} \BibitemShut {NoStop}%
%%CITATION = ARXIV:1306.0533;%%
\bibitem [{\citenamefont {Czech}\ \emph {et~al.}(2014)\citenamefont {Czech},
  \citenamefont {Dong},\ and\ \citenamefont {Sully}}]{Czech:2014wka}%
  \BibitemOpen
  \bibfield  {author} {\bibinfo {author} {\bibfnamefont {Bartlomiej}\
  \bibnamefont {Czech}}, \bibinfo {author} {\bibfnamefont {Xi}~\bibnamefont
  {Dong}}, \ and\ \bibinfo {author} {\bibfnamefont {James}\ \bibnamefont
  {Sully}},\ }\bibfield  {title} {\enquote {\bibinfo {title} {{Holographic
  Reconstruction of General Bulk Surfaces}},}\ }\href {\doibase
  10.1007/JHEP11(2014)015} {\bibfield  {journal} {\bibinfo  {journal} {JHEP}\
  }\textbf {\bibinfo {volume} {11}},\ \bibinfo {pages} {015} (\bibinfo {year}
  {2014})},\ \Eprint {http://arxiv.org/abs/1406.4889} {arXiv:1406.4889
  [hep-th]} \BibitemShut {NoStop}%
%%CITATION = ARXIV:1406.4889;%%
\bibitem [{\citenamefont {Buchel}\ \emph {et~al.}(2010)\citenamefont {Buchel},
  \citenamefont {Escobedo}, \citenamefont {Myers}, \citenamefont {Paulos},
  \citenamefont {Sinha},\ and\ \citenamefont {Smolkin}}]{Buchel:2009sk}%
  \BibitemOpen
  \bibfield  {author} {\bibinfo {author} {\bibfnamefont {Alex}\ \bibnamefont
  {Buchel}}, \bibinfo {author} {\bibfnamefont {Jorge}\ \bibnamefont
  {Escobedo}}, \bibinfo {author} {\bibfnamefont {Robert~C.}\ \bibnamefont
  {Myers}}, \bibinfo {author} {\bibfnamefont {Miguel~F.}\ \bibnamefont
  {Paulos}}, \bibinfo {author} {\bibfnamefont {Aninda}\ \bibnamefont {Sinha}},
  \ and\ \bibinfo {author} {\bibfnamefont {Michael}\ \bibnamefont {Smolkin}},\
  }\bibfield  {title} {\enquote {\bibinfo {title} {{Holographic GB gravity in
  arbitrary dimensions}},}\ }\href {\doibase 10.1007/JHEP03(2010)111}
  {\bibfield  {journal} {\bibinfo  {journal} {JHEP}\ }\textbf {\bibinfo
  {volume} {03}},\ \bibinfo {pages} {111} (\bibinfo {year} {2010})},\ \Eprint
  {http://arxiv.org/abs/0911.4257} {arXiv:0911.4257 [hep-th]} \BibitemShut
  {NoStop}%
%%CITATION = ARXIV:0911.4257;%%
\end{thebibliography}%

\end{document}